\documentclass[aps,reprint,nofootinbib,onecolumn,notitlepage,preprintnumbers]{revtex4-1}
\pdfoutput=1

\usepackage{amsmath}
\usepackage{amssymb}
\usepackage{siunitx}
\usepackage{bbm}
\usepackage{booktabs}
\usepackage{braket}
\usepackage{diagbox}
\usepackage{graphicx}
\usepackage{hyperref}
\usepackage[utf8]{inputenc}
\usepackage{multirow}
\usepackage{placeins}
\usepackage{slashed}
\usepackage[dvipsnames]{xcolor}

\renewcommand{\Re}{\operatorname{Re}}

\newcommand{\eps}{\varepsilon}
\newcommand{\refapp}[1]{appendix~\ref{app:#1}}
\newcommand{\refeq}[1]{eq.~(\ref{eq:#1})}

\newcommand{\reffig}[1]{figure~\ref{fig:#1}}
\newcommand{\refsec}[1]{section~\ref{sec:#1}}
\newcommand{\reftab}[1]{table~\ref{tab:#1}}

\newcommand{\LambdaBar}{\overline{\Lambda}}

\newcommand{\order}[1]{\mathcal{O}\left(#1\right)}

\AtBeginDocument{
  \heavyrulewidth=.08em
  \lightrulewidth=.05em
  \cmidrulewidth=.03em
  \belowrulesep=.65ex
  \belowbottomsep=0pt
  \aboverulesep=.4ex
  \abovetopsep=0pt
  \cmidrulesep=\doublerulesep
  \cmidrulekern=.5em
  \defaultaddspace=.5em
}

\begin{document}

\title{Theory determination of $\boldsymbol{\bar{B}\to D^{(*)}\ell^-\bar\nu}$ form factors at $\boldsymbol{\mathcal{O}(1/m_c^2)}$}

\author{Marzia Bordone}
\email{marzia.bordone@uni-siegen.de}
\affiliation{Universit\"at Siegen, Walter-Flex Stra\ss{}e 3, 57072 Siegen, Germany}

\author{Martin Jung}
\email{martin.jung@unito.it}
\affiliation{Excellence Cluster Universe, Technische Universit\"at M\"unchen, Boltzmannstr. 2, D-85748 Garching, Germany}
\affiliation{Dipartimento di Fisica, Universit\`a di Torino \& INFN, Sezione di Torino, I-10125 Torino, Italy }

\author{Danny van Dyk}
\email{danny.van.dyk@gmail.com}
\affiliation{Technische Universit\"at M\"unchen, James-Franck-Stra\ss{}e 1, 85748 Garching, Germany}

\begin{abstract}
    We carry out an analysis of the full set of ten $\bar{B}\to D^{(*)}$ form factors within the framework of the
    Heavy-Quark Expansion (HQE) to order $\order{\alpha_s,\,1/m_b,\,1/m_c^2}$, both with and without the use of experimental data.
    This becomes possible due to a recent calculation of these form factors at and beyond the maximal physical recoil
    using QCD light-cone sum rules, in combination with constraints from lattice QCD, QCD three-point sum rules and unitarity.
    We find good agreement amongst the various theoretical results, as well as between the theoretical results and the 
    kinematical distributions in $\bar{B}\to D^{(*)}\lbrace e^-,\mu^-\rbrace\bar\nu$ measurements.
    The coefficients entering at the $1/m_c^2$ level are found to be of $\mathcal{O}(1)$, indicating convergence of the HQE.
    The phenomenological implications of our study include an updated exclusive determination of $|V_{cb}|$ in the
    HQE, which is compatible with both the exclusive determination using the BGL parametrization and with
    the inclusive determination.
    We also revisit predictions for the lepton-flavour universality ratios $R_{D^{(*)}}$, the $\tau$ polarization
    observables $P_\tau^{D^{(*)}}$, and the longitudinal polarization fraction $F_L$.
    Posterior samples for the HQE parameters are provided as ancillary files, allowing for their use in subsequent
    studies.
\end{abstract}

\preprint{EOS-2019-02, P3H-19-020, SI-HEP-2019-08, TUM-HEP 1211/19}

\maketitle

\section{Introduction}
The decays $\bar B\to D\ell^-\bar\nu$ and $\bar B\to D^{*}\ell^-\bar\nu$ with $\ell=e,\mu,\tau$ are of great
phenomenological interest for several reasons.  First, the decays with light leptons in the final states are used to
determine the CKM matrix element $|V_{cb}|$ in the Standard Model (SM).  Second, New Physics (NP) scenarios ---
model-independently defined through the means of an Effective Field Theory (EFT) at low energies --- are constrained by
both the light-lepton modes and the ones involving a $\tau$ lepton. Third, the interplay between two heavy quarks
provides a laboratory to study Heavy-Quark Effective Theory (HQET) and the Heavy-Quark Expansion (HQE) of the relevant
hadronic matrix elements, to further our understanding of Quantum Chromodynamics (QCD). Interestingly, there are
presently two tensions between theory predictions and the corresponding experimental measurements: the so-called
\emph{$V_{cb}$ puzzle}, \emph{i.e.} the difference between the value for $|V_{cb}|$ as extracted from inclusive vs.
exclusive modes, and a significant deviation from lepton-flavour universality in ratios of $\tau$ over $\mu$ and $e$
modes \cite{Amhis:2016xyh}.\\

The inference of phenomenological parameters such as $|V_{cb}|$ or the EFT Wilson coefficients from experimental
measurements of branching ratios and kinematical distributions in $\bar B \to D^{(*)}\ell^-\bar\nu$ decays requires
knowledge of the relevant hadronic matrix elements.  The latter are commonly described by a set of ten independent
hadronic form factors, which parametrize the strong-interaction dynamics in these modes as functions of the
four-momentum transfer $q^2$.  The determination of these form factors requires nonperturbative methods, such as
lattice QCD or QCD sum rules.\\

Until recently, the available theoretical calculations were insufficient to fully determine these form factors
independently of experimental data; instead, the form factor shapes and $|V_{cb}|$ were fitted together to the
light-lepton modes. However, this approach requires the assumption of absence of NP in these modes; this does not seem
appropriate, given the anomalies not only in $b\to c\tau\nu$ data, but also in $b\to s\mu^+\mu^-$ modes, since models
accommodating both anomalies commonly also modify the couplings to light leptons in charged-current transitions.
Furthermore, these fits were based on a HQE up to $\mathcal O(1/m_{c,b})$. Recently, the determination of $\bar B\to
D^{(*)}$ form factors has advanced, due to both experimental and theoretical improvements: on the experimental side, the
Belle collaboration has released three measurements of the kinematical distributions of the modes in question, including
correlations between bins \cite{Glattauer:2015teq,Abdesselam:2017kjf,Abdesselam:2018nnh}.  BaBar has performed the first
analysis of the four-fold differential rate \cite{Dey:2019bgc}, however, these data are not yet available in a form that
could be used in our analysis. On the theory side, two lattice determinations of two $\bar{B}\to D$ form factors at
finite recoil became available \cite{Lattice:2015rga,Na:2015kha}.
In addition, a second lattice calculation for one $\bar{B}\to D^*$ form factor at zero recoil was published \cite{Harrison:2017fmw}.
Moreover, a recent light-cone sum rule (LCSR) calculation \cite{Gubernari:2018wyi} provides for the first time
information on all form factors parametrizing matrix elements of the basis of dimension-six operators, including those appearing only
in connection with NP effects. This calculation is complementary to the presently available lattice calculations in that
it is applicable at
$q^2\lesssim 0$, while lattice calculations so far have been carried out at $q^2\gtrsim 8~{\rm GeV}^2$, and only for
a subset of form factors. For those form factors where lattice data are available, the LCSR calculation therefore
acts as an anchor for what would otherwise be an extrapolation of the lattice data based on heavy-quark symmetry
relations. For all other form factors this is the only direct calculation available.

The release of the unfolded Belle data has made it possible, for the first time, to analyze the spectra of $\bar B\to
D^{(*)}\ell^-\bar\nu$ with different approaches for the form factors, while most previous experimental analyses provided their
results in terms of parameters of the \emph{CLN parametrization}, which includes the aforementioned expansion up to
$\mathcal O(1/m_c)$.  The \emph{BGL parametrization}, on the other hand, provides a model-independent parametrization of
form factors based on unitarity and analyticity \cite{Boyd:1995cf}, neither expanding in $1/m_{b,c}$ nor in $\alpha_s$.
Assuming the convergence of the latter expansions, clearly the results obtained from either approach should coincide
asymptotically. Analyses of the recent Belle measurements using the BGL and CLN approaches yielded the following observations:
\begin{itemize}
    \item BGL fits to the unfolded $\bar{B}\to D$ data employing also the recent lattice results work very well
    and yield a value for $|V_{cb}|$ in perfect agreement with the value from inclusive decays \cite{Bigi:2016mdz}. The CLN
    parametrization, while yielding a similar value for $|V_{cb}|$, is not sufficiently flexible to accommodate the
    experimental and lattice data at the same time, indicating the importance of higher-order corrections \cite{Bigi:2016mdz}.
    \item The $\bar{B}\to D$ experimental and lattice data \emph{can} be combined in a HQE framework including consistently the correlations due
    to the parameters in the leading and subleading Isgur-Wise functions, at the cost of introducing partial $1/m_c^2$
    corrections \cite{Bernlochner:2017jka,Jung:2018lfu}.
    \item Comparisons between the BGL and CLN parametrizations using
    the unfolded Belle 2017 data with hadronic tag \cite{Abdesselam:2017kjf} show a surprisingly large difference between
    the values for $|V_{cb}|$, with the value extracted using the BGL parametrization again compatible with the one from
    inclusive decays \cite{Bigi:2017njr,Grinstein:2017nlq,Jaiswal:2017rve}. The central values of such a fit violate
    expectations based on heavy-quark symmetry strongly \cite{Bernlochner:2017xyx}, which is not the case, however, once
    information from the (at the time available) LCSR calculations \cite{Faller:2008tr} is included, at the price of a
    slightly lower increase of $|V_{cb}|$ \cite{Bigi:2017njr,Bigi:2017jbd}.
    \item No such parametrization dependence is
    found when employing the recent untagged Belle results \cite{Abdesselam:2018nnh}, but the value of $|V_{cb}|$ extracted from
    the combined 2017/2018 Belle data remains $\sim 2\sigma$ smaller than the one from inclusive decays
    \cite{Gambino:2019sif}.
\end{itemize}
Given these results, it is fair to say the $V_{cb}$ puzzle is reduced, but
not fully resolved yet. The difficulties in fitting the $\bar{B}\to D$ data and the large differences in the analysis of the
tagged Belle data strongly motivate an analysis of higher-order corrections in the HQE framework.\\

The outline of this article is as follows: in \refsec{FFs} we revisit the heavy-quark symmetry relations for $\bar B\to D^{(*)}$ form factors,
with an emphasis on terms that have been generally omitted so far. In \refsec{FFs:fits} we combine all available theory information on these form
factors, demonstrating the necessity of including additional terms compared to previous treatments. We analyze various scenarios with
different classes of inputs in order to probe their mutual compatibility; we provide the fit results for form factors and quantities of
interest like $R(D^{(*)})$ in the viable scenarios. We also apply our extracted form factors
in fits to the available experimental data in the context of the SM, and show how the inclusion of the additional terms resolves the
deviation in the extracted values of $|V_{cb}|$ in previous fits using the BGL or CLN parametrization. We
summarize our results in \refsec{summary}.

\section{Form factors for $\boldsymbol{\bar{B}\to D^{(*)}}$ transitions}
\label{sec:FFs}

The hadronic matrix elements for $\bar{B}(p)\to D^{(*)}(k)$ semileptonic transitions can be expressed in terms of ten independent
form factors, which are scalar functions of the four-momentum transfer $q^2 \equiv (p - k)^2$. A common basis of form factors arises from
the following definitions: For $\bar{B}\to D$, one commonly defines
\begin{align}
    \label{eq:BPVFF}
    \bra{D(k)} \bar{c} \gamma^\mu b \ket{\bar{B}(p)}
        & = \left[(p + k)^\mu - \frac{M_B^2 - M_D^2}{q^2} q^\mu\right] f^{B\to D}_+(q^2)+ \frac{M_B^2 - M_D^2}{q^2} q^\mu f^{B\to D}_0(q^2)\,,\\
    \label{eq:BPTFF}
    \bra{D(k)} \bar{c} \sigma^{\mu\nu} b \ket{\bar{B}(p)}
        & = \frac{2i}{M_B+M_D}(k^\mu p^\nu-p^\mu k^\nu )f_T(q^2,\mu)\,,
\end{align}
with $\sigma^{\mu\nu}=\frac{i}{2}[\gamma^\mu,\gamma^\nu]$.
In the above, $f_+$ is the vector form factor, $f_T$ is the scale-dependent tensor form factor arising only in NP scenarios
(its definition corresponds to the one in Ref.~\cite{Ball:2004ye}), and $f_0$ doubles as the scalar form factor:
\begin{align}
    \label{eq:BPSFF}
    \bra{D(k)} \bar{c} b \ket{\bar{B}(p)}
        & = \frac{M_B^2 - M_D^2}{m_b - m_c}\, f^{B\to D}_0(q^2)\,.
\end{align}
The matrix elements of the remaining axial and pseudoscalar currents are zero by virtue of QCD conserving parity.\\

For $\bar{B}\to D^*$, one commonly defines
\begin{align}
    \label{eq:BVVFF}
    \bra{D^*(k, \eta)} \bar{c} \gamma^\mu b \ket{\bar{B}(p)}
        & = -\epsilon^{\mu\nu\rho\sigma} \eta^*_{\nu}(k)\, p_\rho\, k_\sigma \frac{2\, V(q^2)}{M_B + M_{D^*}}\,,\\
    \label{eq:BVAFF}
    \bra{D^*(k, \eta)} \bar{c} \gamma^\mu \gamma_5 b \ket{\bar{B}(p)}
        & = i \eta^*_{\nu} \bigg\lbrace
                2 M_{D^*} A_0(q^2) \frac{q^\mu q^\nu}{q^2}+ 16\frac{M_B  M_{D^*}^2}{\lambda} A_{12} \left[2 p^\mu q^\nu - \frac{M_B^2 - M_{D^*}^2 + q^2}{q^2} q^\mu q^\nu\right]\\
        & \phantom{=} \nonumber
                + (M_B + M_{D^*})\, A_1(q^2) \left[g^{\mu\nu} + \frac{2(M_B^2 + M_{D^*}^2 - q^2)}{\lambda} q^\mu q^\nu
                - \frac{2(M_B^2 - M_{D^*}^2 - q^2)}{\lambda} p^\mu q^\nu\right]\bigg\rbrace\,,\\
    \label{eq:BVTFF}
    \bra{D^*(k, \eta)} \bar{c} \sigma^{\mu\nu} b \ket{\bar{B}(p)}
        & = i
        \eta^*_{\alpha}\epsilon^{\mu\nu}\!{}_{\rho\sigma}\left\{-\left[\left((p+k)^\rho-\frac{M_B^2-M_{D^*}^2}{q^2}q^\rho\right)g^{\alpha\sigma}+
        \frac{2}{q^2} p^\alpha p^\rho k^\sigma\right]T_1(q^2)\right.\\
        &  \qquad\qquad\quad\,\,\,-\left.\left(\frac{2}{q^2}p^\alpha p^\rho k^\sigma-\frac{M_B^2-M_{D^*}^2}{q^2}q^\rho
        g^{\alpha\sigma}\right) T_2(q^2)+\frac{2}{M_B^2-M_{D^*}^2}p^\alpha p^\rho k^\sigma T_3(q^2)\right\}\,.\nonumber
\end{align}
where $\eta$ denotes the $D^*$ polarization vector, $V$ the vector form factor, and $A_{1,12}$ are the axial form factors.
Note that the relative sign
between our \refeq{BVVFF} and the decomposition in ref.~\cite{Ball:2004rg} arises from the different definition
of the Levi-Civita tensor: we use $\varepsilon^{0123} = +1$.
Moreover, in the decomposition above $A_{12}$ correspond to longitudinal polarizations of the emitted virtual $W$, which
is more convenient (e.g. when inferring form factors  from lattice QCD) than parametrizations involving the form factor $A_2$, see
e.g. \cite{Ball:2004rg}.
The function $A_0$ doubles as the pseudo-scalar form factor,
\begin{align}
    \label{eq:BVPFF}
    \bra{D^*(k, \eta)} \bar{c} \gamma_5 b \ket{\bar B(p)}
        & = -2i M_{D^*}\,\frac{\eta^* \cdot q}{m_b + m_c} A_0\,,
\end{align}
whereas the matrix element of the scalar current vanishes by virtue of QCD conserving parity.\\

Exact relations at $q^2 = 0$ between some of the form factors ensure the absence of unphysical singularities in \refeq{BPVFF} and \refeq{BVAFF}.
These relations read:
\begin{equation}
    \label{eq:relations-at-q2-zero}
    \begin{aligned}
        f_+(q^2 = 0)
            & = f_0(q^2 = 0)\,,                                                                                  \\
        A_0(q^2 = 0)
            & = \frac{M_B + M_{D^*}}{2 M_{D^*}}\,A_1(q^2 = 0) - \frac{M_B - M_{D^*}}{2 M_{D^*}}\,A_2(q^2 = 0)\,.
    \end{aligned}
\end{equation}
A further exact relation arises due to algebraic identities involving the Lorentz structures $\sigma^{\mu\nu}$ and
$\sigma^{\mu\nu}\gamma_5$~\cite{Ball:2004rg}:
\begin{equation}
    T_1(0) = T_2(0)\,.
\end{equation}
Further approximate relations arise from the HQE of the hadronic matrix elements.
These relations, the parametric models involved, and theoretical inputs needed for the
subsequent statistical analyses are the subject of the remainder of this section.

\subsection{Heavy-Quark Expansion and models}
\label{sec:FFs:HQE}

The combination of heavy-quark spin symmetry and heavy-quark flavour symmetry permits to relate
$\bar{B}^{(*)}(v) \to D^{(*)}(v')$ matrix elements with each other in a simultaneous expansion in the strong coupling $\alpha_s$ and 
the inverse pole masses $1/m_Q$, where $Q=b,c$ is the quark flavour.
The coefficients of this HQE -- up to kinematical and combinatorial factors -- are the
Isgur-Wise functions, which depend exclusively on the recoil parameter $w \equiv v\cdot v'$. For convenience, the
expansion is commonly expressed in terms of dimensionless quantities $\eps_Q \equiv \LambdaBar / 2 m_Q$, where $\LambdaBar$
arises in the HQE of the heavy meson masses.\\

We begin by adopting the power counting $\eps_b\sim\eps_c^2\sim\alpha_s/\pi\sim\eps^2$ for the HQE. Consequently, when expanding up to $\mathcal O(\eps^2)$, we need to account for all leading-order radiative and subleading-power corrections, as well as partial subsubleading-power corrections.
Higher powers in our expansion or mixed terms are assumed to be negligible. The HQE is well known, and we follow ref.~\cite{Bernlochner:2017jka} closely in our analysis.
By virtue of our power counting, any form factor $F$ discussed in \refsec{FFs} can be expressed in terms of
ten independent functions: $\xi$, the leading Isgur-Wise (IW) function;
$\chi_{2,3}$ and $\eta$, the subleading IW functions; and $\ell_{1,\dots,6}$, the subsubleading IW functions at order $\eps_c^2$
as introduced in ref.~\cite{Falk:1992wt}; see also \refapp{FF:HQE} for more details.
Each of these functions depends on the recoil parameter $w$. In the complex half plane $\Re{w} \geq 1$,
the form factors, the HQET Wilson coefficients, and the IW functions are free of singularities due to QCD dynamics. Singularities of
kinematical origin can always be removed by redefining the form factors. Consequently, for $f$ being
any of the ten IW functions considered here, we expand it around $w = 1$:\footnote{%
    Note that the form factors are also commonly written as \cite{Caprini:1997mu} 
    $\tilde f(w) = \tilde f(w_0)(1-\rho^2(w-w_0)+c(w-w_0)^2+d(w-w_0)^3+\ldots)$,
    such that $f'(w_0)=-\tilde f(w_0)\rho^2,f''=2!\tilde f(w_0)c$, etc.
}
\begin{equation}
\begin{aligned}
    \label{eq:def:w-expansion}
    f(w)
        & = \sum_{k=0}^K \frac{f^{(k)}}{k!}\, \left(w - 1\right)^k\,.
\end{aligned}
\end{equation}
Following ref.~\cite{Boyd:1995cf,Caprini:1997mu}, we can further trade the variable $w$ for $z(w)$, which correctly captures the analytic
properties of the matrix elements, \emph{i.e.} it develops a branch cut corresponding to the $B^{(*)} D^{(*)}$ pair production at $w \leq -1$.
While $z(w)$ is a small expansion parameter in the semileptonic phase space, absent further modifications to the form factors as discussed
in ref.~\cite{Boyd:1995cf} we cannot generally expect small coefficients in an expansion in $z$. We proceed to expand each monomial
$(w-1)^k$ in \refeq{def:w-expansion} around $z(1)$, where the maximum order in $z - z(1)$ depends on our concrete parameter models discussed
later. In this way, we keep the benefits inherent to parametrizing in $z$, while conserving at the same time the physical meaning
of the fit parameters $f^{(n)}(1)$ as derivatives of the IW functions at the zero-recoil point. In this setup, we follow
ref.~\cite{Jung:2018lfu} closely.\\

Both HQET and the HQE of the heavy meson masses provide us with some
information on the parameters arising in the HQE of the hadronic matrix
elements at hand. The remaining ones need inferring from theoretical or
experimental inputs. For our statistical analyses we define fit models,
which vary only in our choice of the order to which the different Isgur-Wise functions are expanded in $z$.
All our models include all ten Isgur-Wise functions above;
the expansion up to $1/m_c^2$ is not only preferable from the point of view of precision, but, as mentioned in the
introduction, necessary, given the available lattice data at $w=1$.  Employing the recent LCSR results
\cite{Gubernari:2018wyi} allows to include \emph{all} subsubleading IW functions, which is an improvement compared to
ref.~\cite{Jung:2018lfu}, where only the functions $\ell_{1,2}$ could be included.
The models used in this work are denoted as $k/l/m$, where the numbers $k$, $l$ and $m$ have the following meaning:
\begin{itemize}
    \item[$k:$] the order to which the leading IW function is expanded in $z$ around $z=0$;
    \item[$l:$] the order to which the subleading IW functions are expanded; and
    \item[$m:$] the order to which the subsubleading IW functions are expanded.
\end{itemize}
We keep all purely kinematical powers of $w$, \emph{i.e.}~terms that arise when relating the form factors to the IW functions
in the HQE. Within the scope of this work we will discuss the models $2/1/0$ and $3/2/1$.\\

We emphasize that increasing the maximum order in the $z$ expansion from one fit model to another can always be
expressed in terms of a non-zero shift to the new parameter appearing in the higher-order term. As an example, consider
increasing the order of the $z$ expansion from a $2/l/m$ to a $3/l/m$ model. The $2/l/m$ models
can be recovered in the $3/l/m$ model by assigning $\xi'''(1) = -\xi''(1) / 2 - 3 \xi'(1) / 64$.

\subsection{Theory constraints}
\label{sec:FFs:theory}

With the definition of the models at hand we proceed to the available theoretical calculations of the
hadronic matrix elements as well as theoretical bounds on the parameter space derived from dispersion relations \cite{Boyd:1995cf,Caprini:1997mu}.
The individual pieces of theory information entering the likelihood are:
\begin{description}
    \item[Lattice] For $\bar{B}\to D$ the HPQCD and FNAL/MILC collaborations have, independently from each other, determined the
        vector form factor $f_+$ and the scalar form factor $f_0$ at several values of the recoil parameter $w \geq 1$.
        We use correlated pseudo data points from both studies: seven from \cite{Lattice:2015rga} and five from \cite{Na:2015kha}.
        Note that at $w = w_{\text{max},D}$ the form factors fulfill an equation of motion that reduces the number of
        observations per study.\\[\smallskipamount]
        For $\bar{B}\to D^*$ the HPQCD and FNAL/MILC collaborations have independently determined the form
        factor $h_{A_1}$ at $w = 1$ \cite{Bailey:2014tva,Harrison:2017fmw},
        averaged by FLAG to $h_{A_1}(w=1)=0.904\pm 0.012$ \cite{Aoki:2016frl}.
    \item[QCDSR] The subleading IW functions $\chi_{2,3}$ and $\eta$ have been studied within three-point QCD Sum Rules
        \cite{Neubert:1992wq,Neubert:1992pn,Ligeti:1993hw}. These sum rules have been used to infer the normalization and slope of
        the subleading IW functions at $w=1$, yielding five observations in total.
    \item[LCSR] At $w \gtrsim 1.5$ the $\bar{B}\to D^{(*)}$ form factor can be accessed using LCSRs with $B$-meson Light-Cone Distribution Amplitudes (LCDAs)
        \cite{Faller:2008tr}. These results have been superseded by an updated analysis \cite{Gubernari:2018wyi}, which includes
        for the first time all two-particle and three-particle LCDAs in a consistent twist-expansion up to twist 4 \cite{Braun:2017liq}.
        Moreover, the recent analysis provides for the first time information about the shape of the complete set of $\bar{B}\to D^*$
        form factors at four phase space points, albeit with two caveats: the form factors are available only at $w \geq w_\text{max}\simeq 1.5$
        and the $\bar{B}\to D$ form factor $f_T$ could not be extracted as part of the same approach as the other form factors.
        The first point requires attention in the context of the $z$ expansion, since it increases the maximal value of $|z|$ to values
        larger than encountered in other studies.
        The second point dissuades us from including $f_T$ in the analysis; instead, we choose to predict $f_T$ within our approach and
        compare it with the prediction of ref.~\cite{Gubernari:2018wyi}.
        Following the introductory discussion concerning exact relations between some of the form factors, we arrive at a total number
        of 33 observations.
\end{description}
Beyond the likelihood, we include further information on the hadronic matrix elements. This additional
information is expressed as so-called unitarity bounds~\cite{Boyd:1995cf,Caprini:1997mu}. In the context of the HQE, it is convenient to
adopt the approach of ref.~\cite{Caprini:1997mu}. We consider the bounds for the currents $J_{0^+} \equiv \bar{c}b$,
$J_{0^-} \equiv \bar{c}\gamma_5 b$, $J_{1^-}^\mu \equiv \bar{c}\gamma^\mu b$, and $J_{1^+}^\mu \equiv \bar{c}\gamma^\mu\gamma_5 b$.
For all currents $J_{L^P}$ we derive the bounds in terms of the full set of ten independent IW functions present in our
models. The results of the perturbative OPE calculations for the bounds are denoted as $\chi_{L^P}$ and $\tilde{\chi}_{L^P}$, where
the tilde indicates subtraction of known one-body contributions~\cite{Caprini:1997mu}. Updated values for these four quantities
have been recently provided in ref.~\cite{Bigi:2016mdz,Bigi:2017njr,Bigi:2017jbd} based on ref.~\cite{Grigo:2012ji}.
The general problem of how to include positivity bounds in a Bayesian fit and our approach to solve it is discussed in
\refapp{bounds}.\\

In the construction of the unitarity bounds, a choice must be made to which order $n$ in $z$ the bound is formulated. Using BGL-like form
factors, this coincides with the order to which the form factors are expanded. However, the treatment of the unitarity bounds in the
context of the HQE is non-trivial. The reason for their complexity arises from the simultaneous expansion in $1/m_Q$,
$\alpha_s$, and~$z$.
As indicated above, we expand the IW functions on different levels in the $1/m_Q$ expansion to different orders in $z$,
according to their \emph{combined} power-counting, \emph{i.e.}, a $z^3$ contribution might be relevant for the form
factors when entering the leading IW function, while such a term in a subsubleading IW function is expected to be
negligible. Hence we choose generally $k\geq l\geq m$. However, in a BGL setup the unitarity bounds are written as
quadratic forms of the BGL coefficients \emph{without} explicit factors of $z$. Therefore the relative importance of
higher-order contributions in $z$ is larger in these bounds than in the form factors themselves.
Consequently, the treatment of these higher-order contributions is important. Specifically, $1/m_Q^2$ contributions are
only fully included for $n\leq m$, $1/m_Q$ contributions for $n\leq l$ and leading-order contributions for $n\leq k$.
Since particularly $1/m_c$ contributions can be large, and the terms in the $1/m_Q$ expansion are not necessarily positive,
the order $n$ for the unitarity bounds should be chosen to be at most $n=l$, with $l\geq 1$.\\

The combination of the described constraints allows to include higher-order contributions in the HQE for the full set of form factors.
Within our determination of these contributions we pose the following questions:
\begin{itemize}
    \item Are the various theoretical constraints mutually compatible in the context of the HQE? If yes, what is the minimal $k/l/m$ model
    that achieves a good description?
    \item In case of a successful combined fit, what are the phenomenological consequences with respect to $|V_{cb}|$ and predictions for $B\to D^{(*)}\tau\nu$ observables?
\end{itemize}

\section{Statistical analyses}
\label{sec:FFs:fits}

The numerical and statistical results presented in the following have been obtained by means of two
completely independent implementations. One of these is publicly available as part of the \texttt{EOS}
software \cite{EOS}, which has also been used to prepare all of the following numerical results
and plots. The posterior samples used to produce these results and plots are available as ancillary files
\cite{EOS-DATA-2019-01}.

\subsection{Fits to only theory constraints and SM predictions}

The minimal model fulfilling all criteria laid out in the previous section is the $1/1/0$ model. We find this model to
provide a bad fit to the available theory constraints, with $\chi^2\sim 560$ in the best-fit point for $39$ degrees of
freedom (dof). We therefore proceed to fit the theory constraints with the $2/1/0$ model, which yields an excellent fit
with $\chi^2=22.87$ for $38$~dof. This model therefore represents the minimal \emph{viable} fit model.\footnote{%
    Abandoning the requirement $l\geq 1$ leads to another model with an excellent fit, $2/0/0$, whose best-fit point
    essentially coincides with the one of the $2/1/0$ model.
}
Following the discussion in the previous section, it is important to account for systematic uncertainties inherent to
the HQE by increasing the order of the $z$ expansions by one. The corresponding model, $3/2/1$, reduces the $\chi^2$ in the
best-fit point by $\sim 13$, at the expense of $10$ additional parameters. Details for these fits are given in
table~\ref{tab:gof}.\\

\begin{figure}[t!]
    \hspace*{-.06\textwidth}
    \begin{tabular}{ccc}
        \includegraphics[width=.35\textwidth]{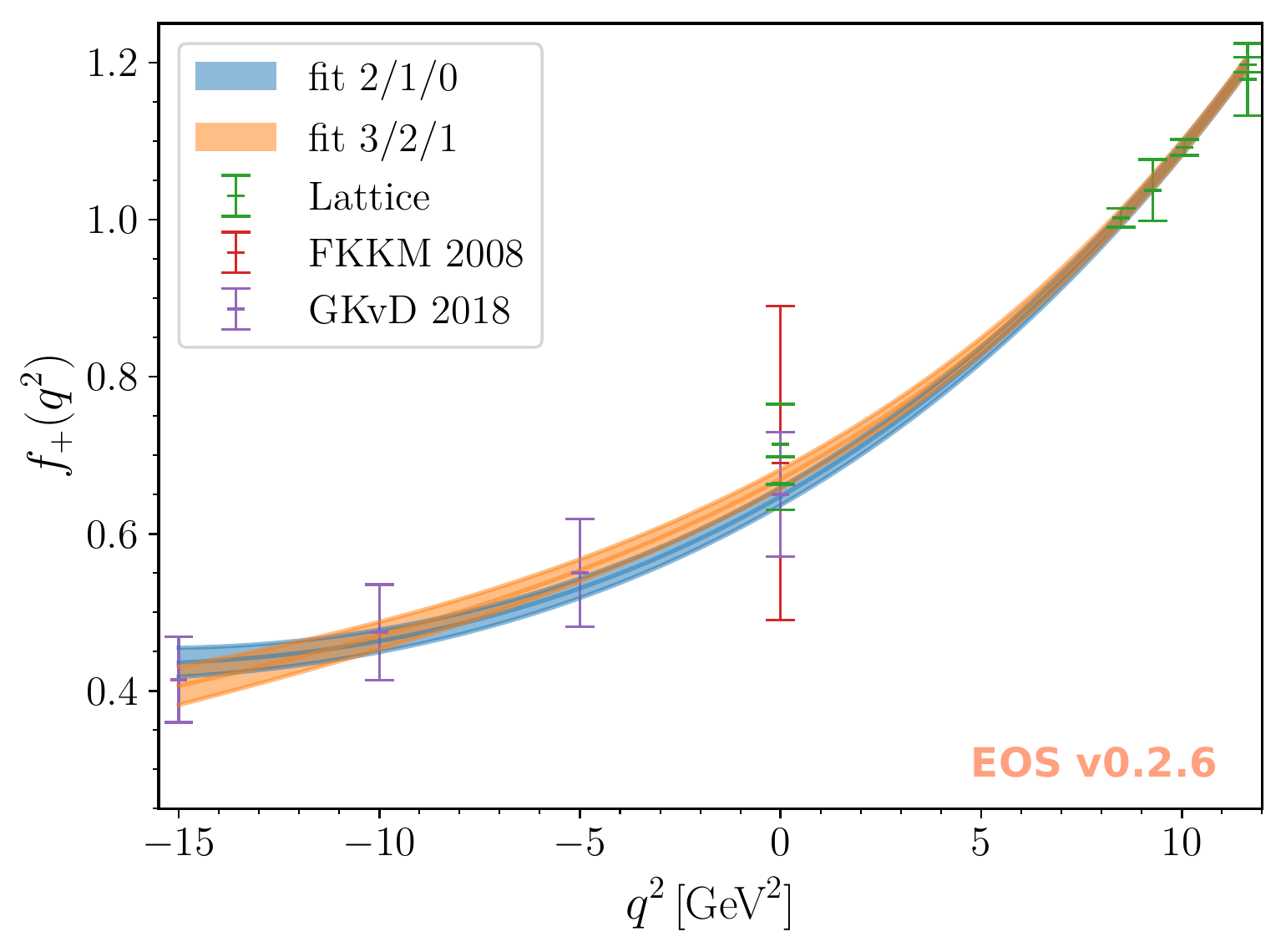}      &
        \includegraphics[width=.35\textwidth]{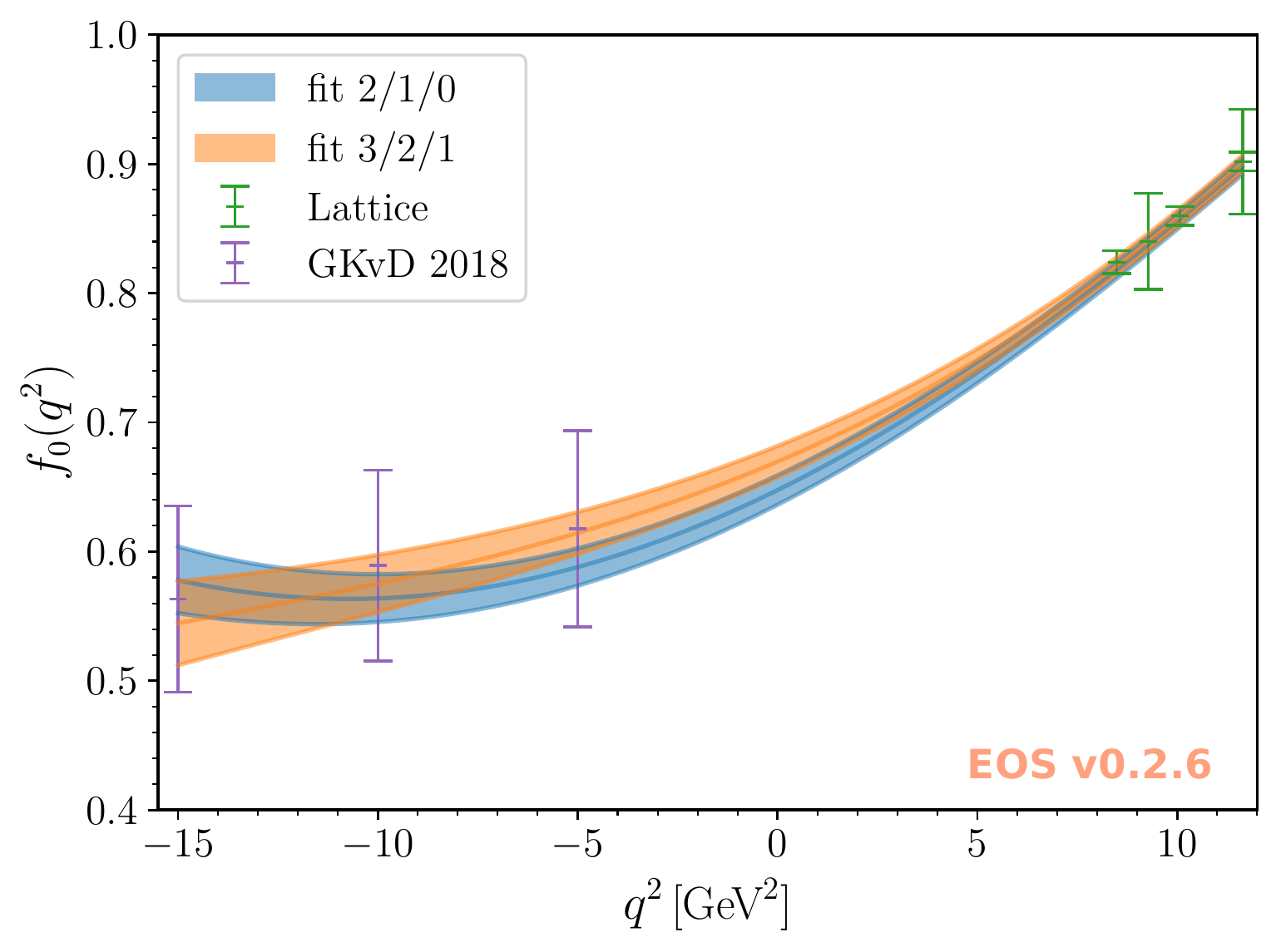}      &
        \includegraphics[width=.35\textwidth]{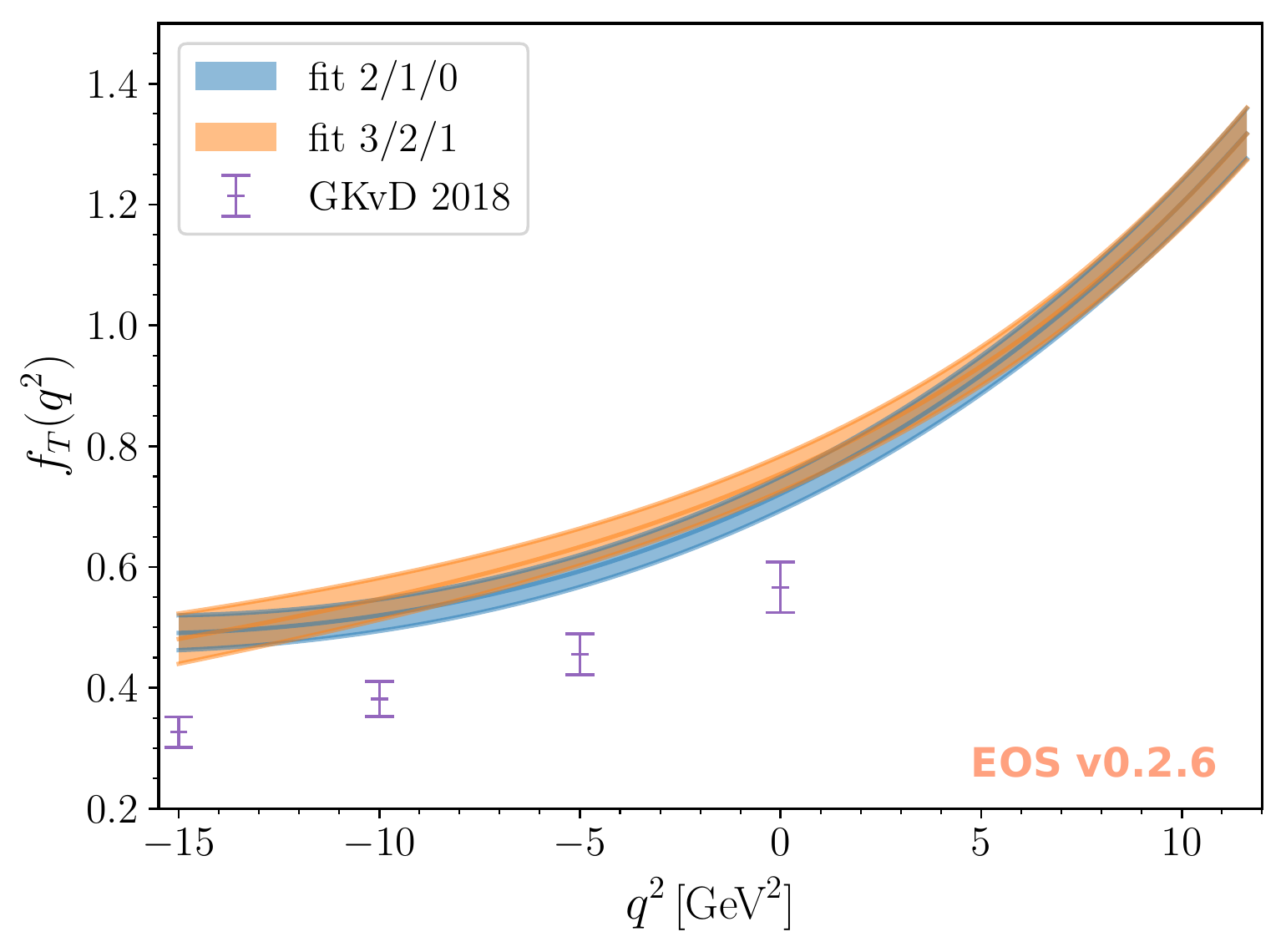}      \\[1.25em]
        \includegraphics[width=.35\textwidth]{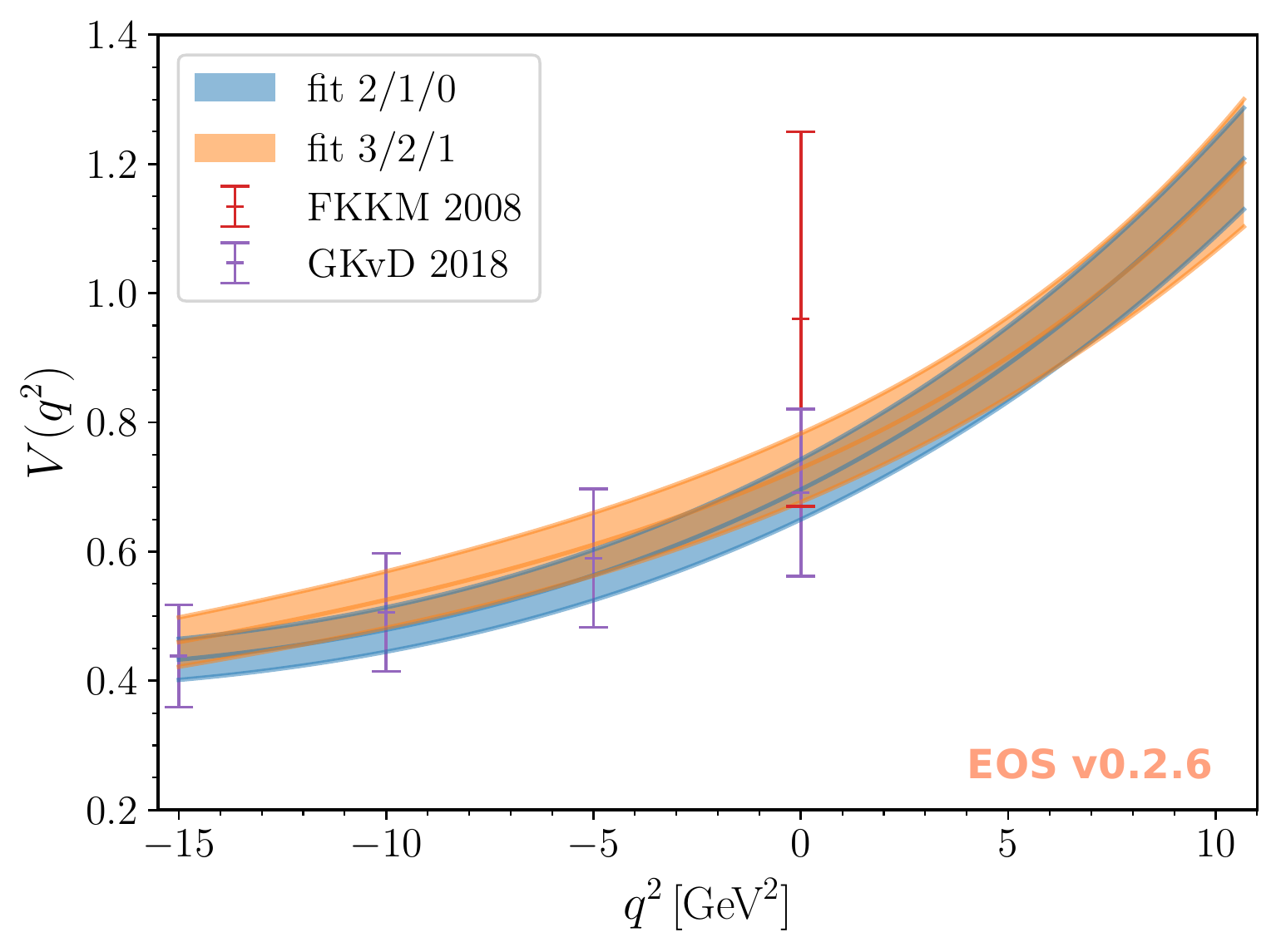}   &
        \includegraphics[width=.35\textwidth]{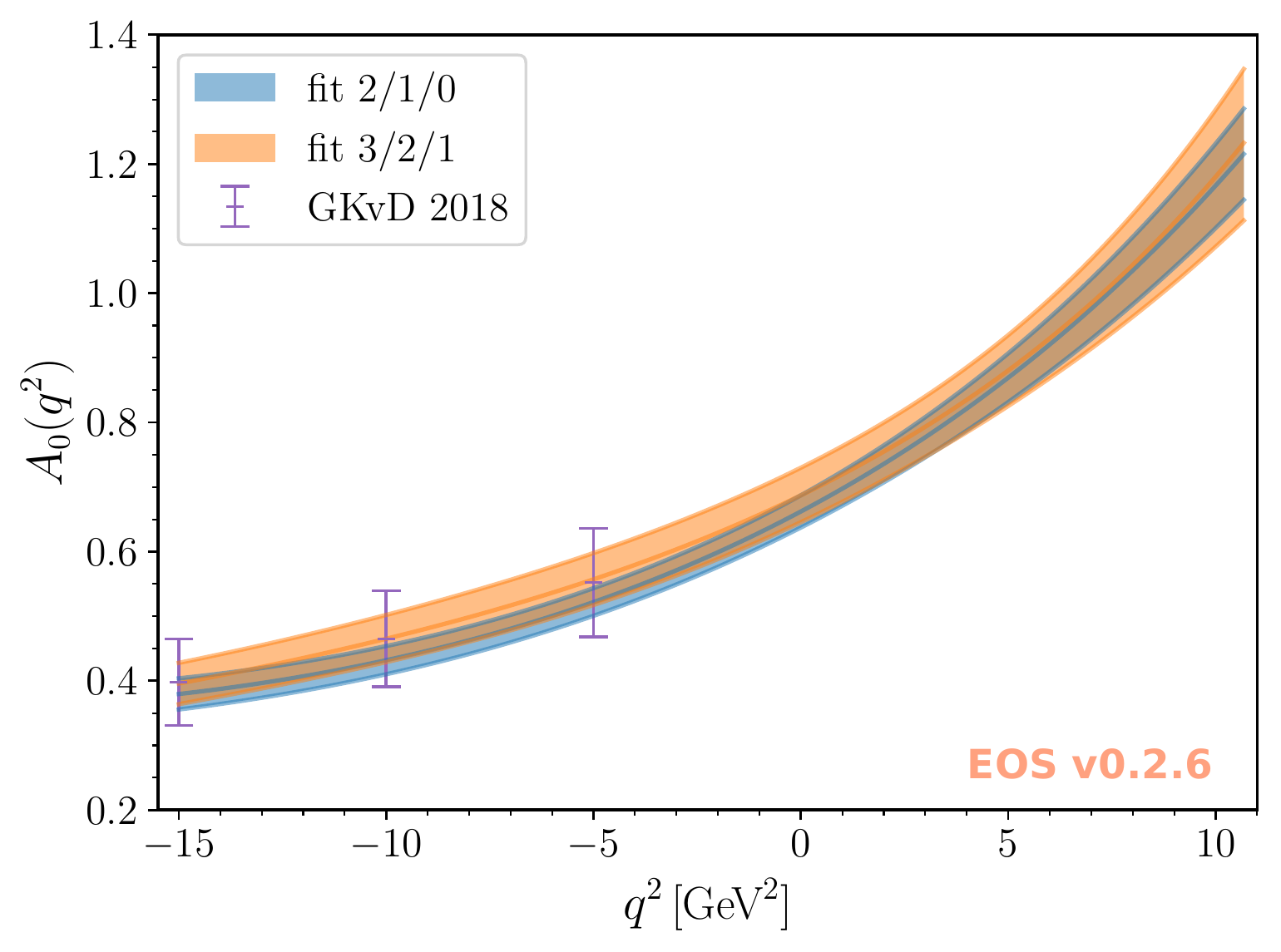}  &
        \includegraphics[width=.35\textwidth]{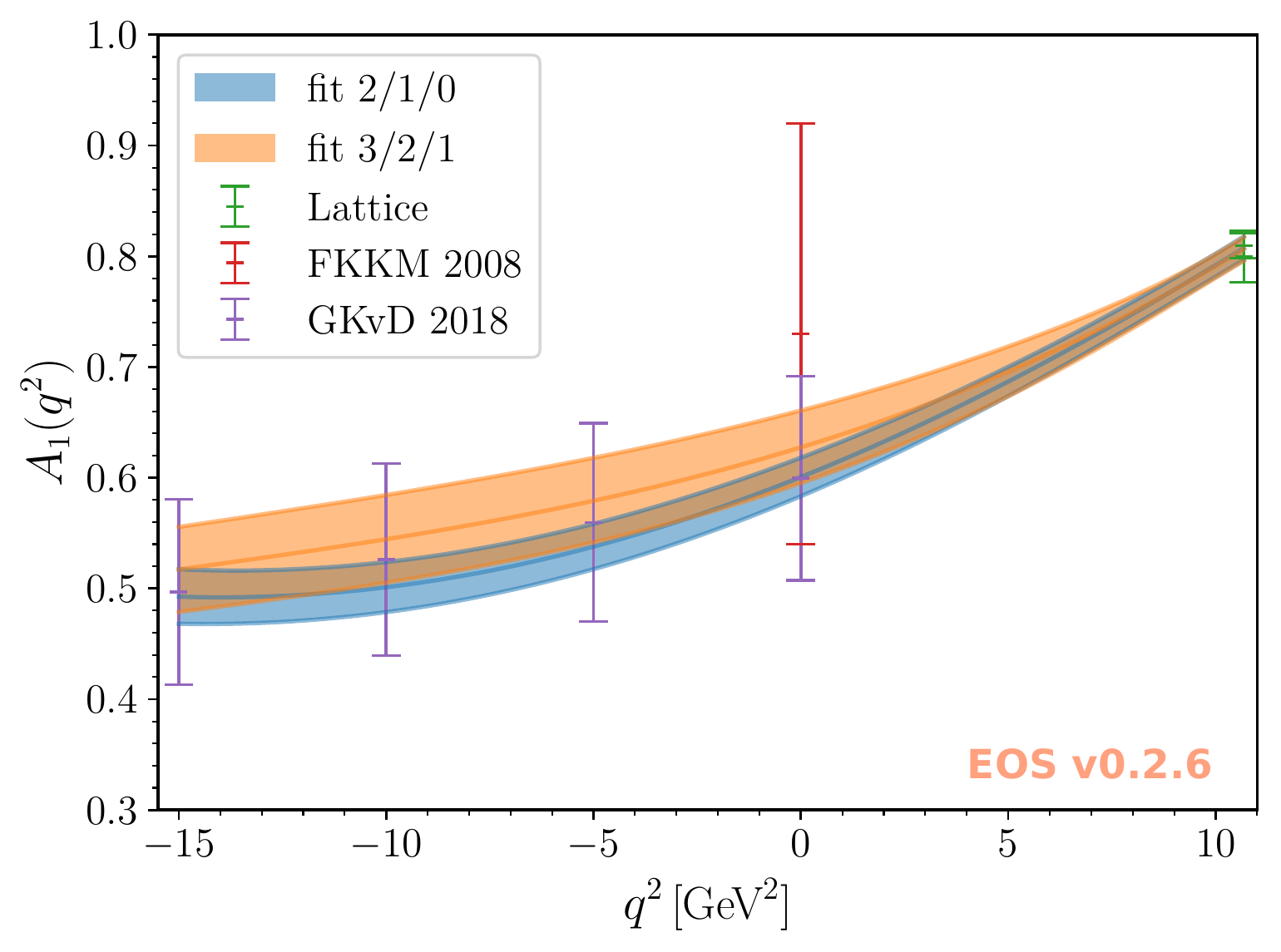}  \\[1.25em]
        \includegraphics[width=.35\textwidth]{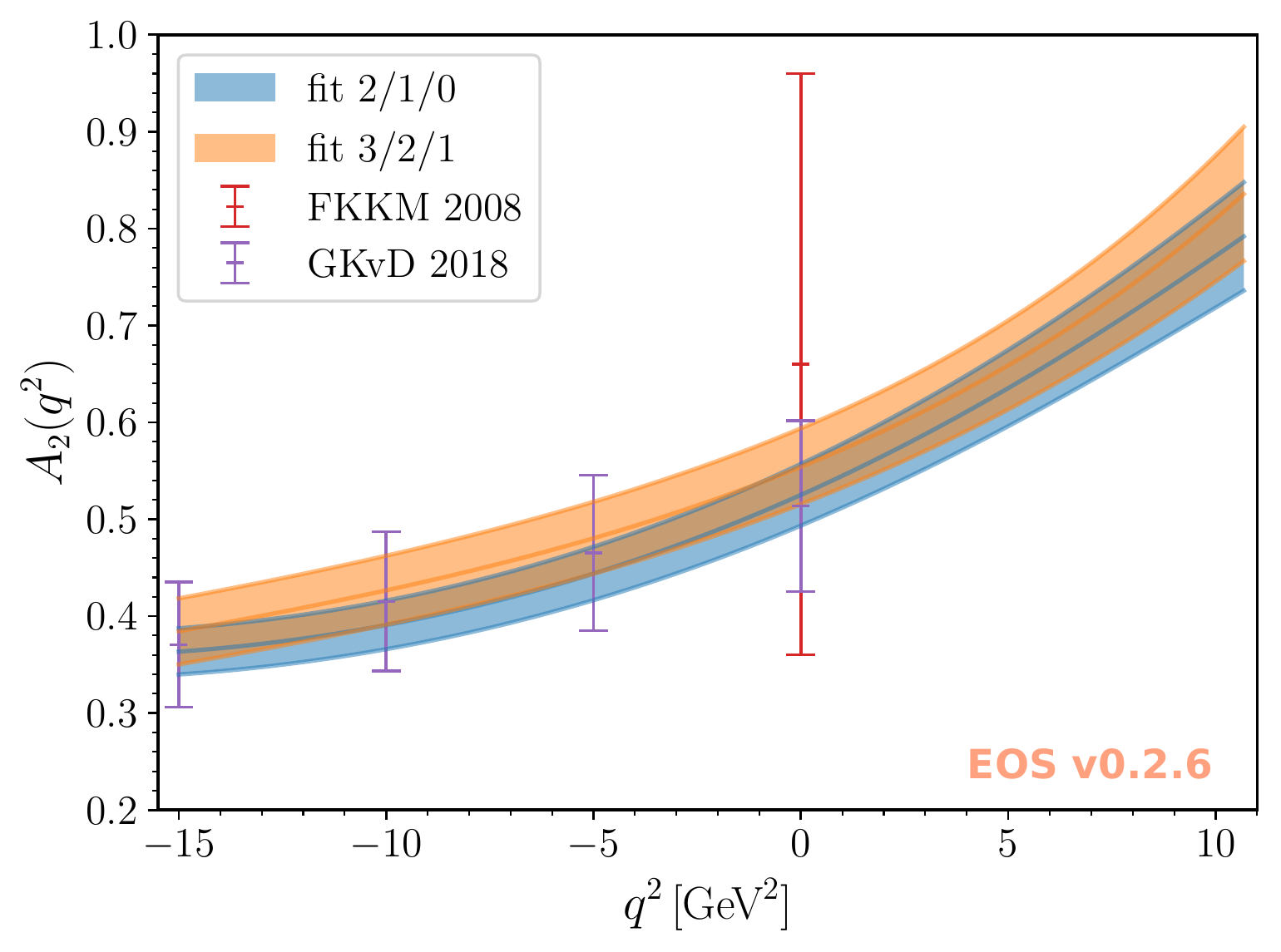}  &
        \includegraphics[width=.35\textwidth]{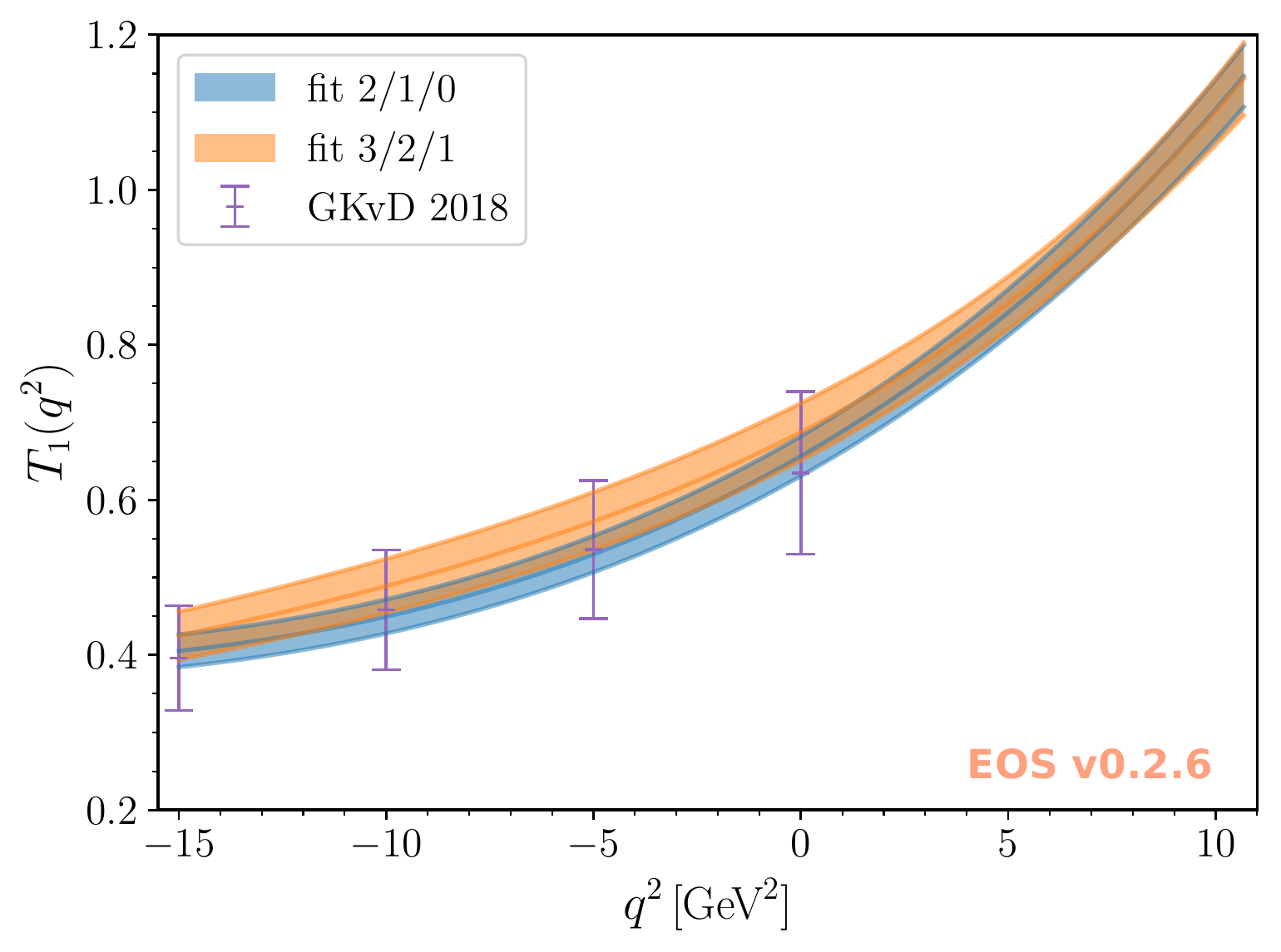}  &
        \includegraphics[width=.35\textwidth]{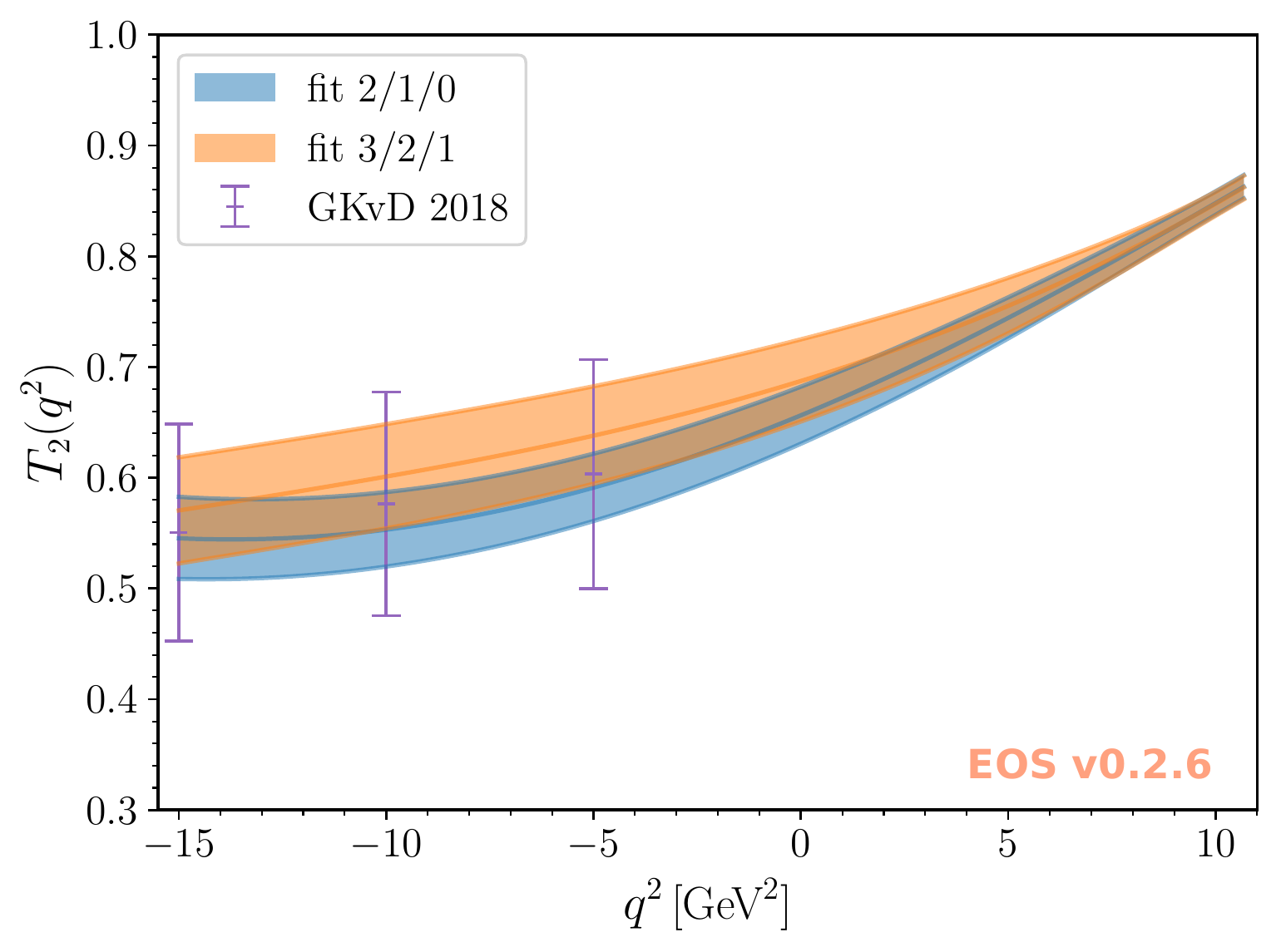}  \\[1.25em]
        \includegraphics[width=.35\textwidth]{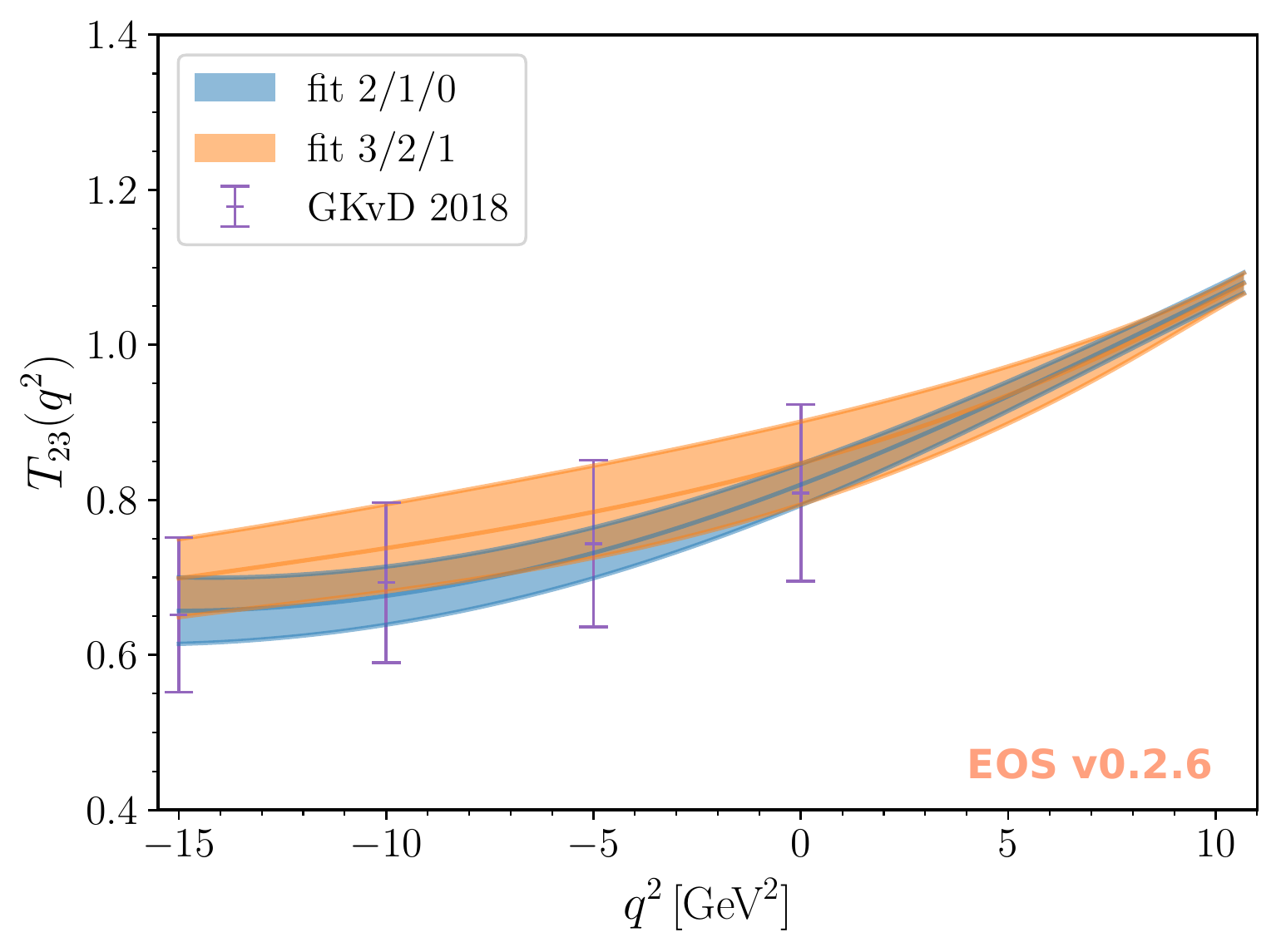}
    \end{tabular}
    \caption{
        The full set of $\bar{B}\to D^{(*)}$ form factors as a function of $q^2$ are used to showcase
        our results for the nominal fit model $3/2/1$ (orange lines and areas), in comparison to the minimal
        viable fit model $2/1/0$ (light blue lines and areas).
        For both models we show the central values and $68\%$ probability envelopes from posterior-predictive
        distributions of the respective fits.
        The lattice constraints used in the fits are shown as green data points.
        The LCSR constraints used in the fits are shown as purple data points.
        The superseded LCSR results \emph{not} used in the fits are shown as red data points
        for comparison, only.
    }
    \label{fig:formfactors}
\end{figure}

Using samples of the posteriors of the fits to both models, we produce posterior predictive distributions for all $\bar
B\to D^{(*)}$ form factors, including $f_T$. The median values and $68\%$ probability envelopes for each
form factor are shown in \reffig{formfactors}, together with data points illustrating the theory constraints where
applicable. We make the following observations:
\begin{itemize}
    \item As expected, the uncertainty bands are systematically broader in the $3/2/1$ model.
    \item For the form factors $f_0, A_1$ and~$T_2$ we observe that model $2/1/0$ produces a local minimum for $-15~{\rm
    GeV}^2\leq q^2\leq 0$, where the LCSR constraints are available. This does not conform to the usual expectation in a
    dispersive picture: Far below the production threshold and sub-threshold poles it should be possible to approximate the
    dispersive integrals for the form factors with a single effective pole, leading to a monotonically falling form factor
    with decreasing $q^2$.
    \item Neither of the two models is able to simultaneously fit all the nominal theory constraints plus the LCSR
    constraints on the $\bar B\to D$ form factor $f_T$. This is not surprising, given the different framework used for its
    prediction compared to the other form factors, as discussed in ref.~\cite{Gubernari:2018wyi}.
\end{itemize}
In addition, we use the posterior samples for the $3/2/1$ fit model to produce posterior-predictive distributions in the
SM for the LFU ratios $R_D$ and $R_{D^*}$, the $\tau$ polarizations $P_\tau^{D^{(*)}}$ in $\bar{B} \to D^{(*)}\tau^-\bar\nu$
decays, and the longitudinal polarisation fraction $F_L$ in $\bar{B}\to D^*\tau^-\bar\nu$ decays. We obtain
\begin{equation}
\begin{aligned}\label{eq:bctaunupred}
    R_{D}
        & = 0.298 \pm 0.003\,, &
    R_{D^*}
        & = 0.247 \pm 0.006\,, \\
    P_\tau^{D}
        & = 0.321 \pm 0.003\,, &
   -P_\tau^{D^*}
        & = 0.488 \pm 0.018\,, \\
        &                      &
    F_L
        & = 0.470 \pm 0.012\,.
\end{aligned}
\end{equation}

\begin{table}[t]
    \renewcommand{\arraystretch}{1.1}
    \begin{tabular}{r c c c c c c c}
        \toprule
        ~                              &
            ~                          &
            $2/1/0$                    &
            $3/2/1$                    &
            $3/2/1$                    &
            $3/2/1$                    &
            $3/2/1$                    \\
        likelihood                     &
            \diagbox{\#obs}{\#par}     &
            $13$                       &
            $23$                       &
            $23$                       &
            $23$                       &
            $23$                       \\
        \midrule
        lattice($D$) &
            $12$     &
            $11.15$  &
            $ 7.06$  &
            $ 7.29$  &
            $ 7.29$  &
            $ 7.36$  \\
        lattice($D^*$) &
            $1$      &
            $ 0.00$  &
            $ 0.01$  &
            $ 0.00$  &
            $ 0.01$  &
            $ 0.00$  \\
        QCDSR        &
            $5$      &
            $ 4.58$  &
            $ 0.04$  &
            $ 0.04$  &
            $ 0.01$  &
            $ 0.02$  \\
        LCSR         &
            $33$     &
            $ 7.14$  &
            $ 2.79$  &
            $ 3.23$  &
            $ 3.14$  &
            $ 2.98$  \\
        \midrule
        $\bar{B}\to D \{e^-,\mu^-\}\bar\nu$ &
            $(9)$    &
            ---      &
            ---      &
            ---      &
            ---      &
            $ 6.75$  \\
        $\bar{B}\to D^* \{e^-,\mu^-\}\bar\nu$ 2017 &
            $(9)$    &
            ---      &
            ---      &
            $ 6.95$  &
            ---      &
            $ 8.04$  \\
        $\bar{B}\to D^* \{e^-,\mu^-\}\bar\nu$ 2018 &
            $(9)$    &
            ---      &
            ---      &
            ---      &
            $ 4.42$  &
            $ 4.84$  \\
        \midrule
        \multirow{3}{*}{total} &
            $51$               &
            $22.87$            &
            $ 9.91$            &
            ---                &
            ---                &
            ---                \\
                               &
            ($60$)             &
            ---                &
            ---                &
            $17.51$            &
            $14.88$            &
            ---                \\
                               &
            ($78$)             &
            ---                &
            ---                &
            ---                &
            ---                &
            $30.00$            \\
        \bottomrule
    \end{tabular}
    \renewcommand{\arraystretch}{1}
    \caption{%
        Summary of the goodness of fit at the best-fit point for all combinations of fit models and datasets.
        The largest single pull arises from the QCDSR constraint on $\chi_2'(1)$
        with $\chi^2 \simeq 4$.
    }
    \label{tab:gof}
\end{table}

We also produce posterior-predictive distributions for the $\bar{B}\to D^{(*)}\{e^-,\mu^-\}\bar\nu$ branching ratios for both of
our fit models. Their summaries in form of mean value, standard deviations and correlations are collected in \reftab{BRs+Vcb}.

\subsection{Challenging measurements and extraction of $|V_{cb}|$}

We apply the form factors obtained in the previous subsection to the available experimental information to perform 
phenomenological studies with high accuracy. Specifically, we confront our predictions with the measured spectral
information and extract $|V_{cb}|$, assuming the SM. Our extraction of $|V_{cb}|$ to subsubleading power in the HQE
is the first of its kind.\\

The publicly available experimental results are $\bar{B}\to D^{(*)}\ell^-\bar\nu$ kinematical distributions
published by the Belle collaboration \cite{Glattauer:2015teq,Abdesselam:2017kjf,Abdesselam:2018nnh} and the world
averages for the branching fractions \cite{Amhis:2016xyh}. The $\bar{B}\to D\ell^-\bar\nu$ distribution $P_D(w)$ from
ref.~\cite{Glattauer:2015teq}, and the four $\bar{B}\to D^*\ell^-\bar\nu$ distributions $P_{D^*}(w)$, $P_{D^*}(\chi)$,
$P_{D^*}(\cos \theta_{D^*})$, and $P(\cos\theta_\ell)$ from ref.~\cite{Abdesselam:2017kjf}\footnote{%
    Note that these results are still preliminary and a new analysis of the data is ongoing.
} are
unfolded of detector effects by the Belle collaboration. The data presented in ref.~\cite{Abdesselam:2018nnh} are still
folded, and the necessary information for the unfolding process is provided in the publication.\\

\begin{figure}[htp]
    \begin{tabular}{cc}
        \includegraphics[width=.49\textwidth]{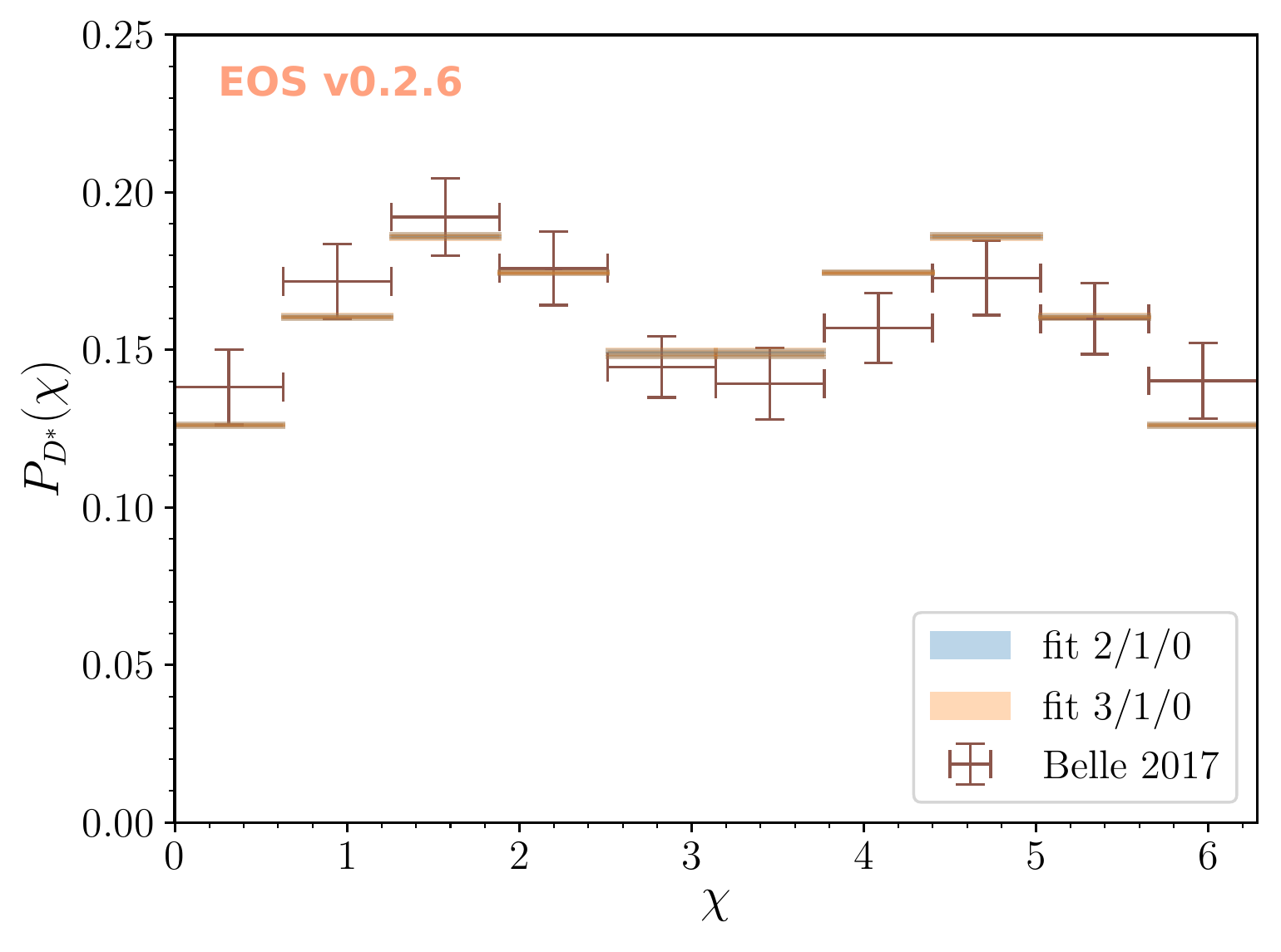}  &
        \includegraphics[width=.49\textwidth]{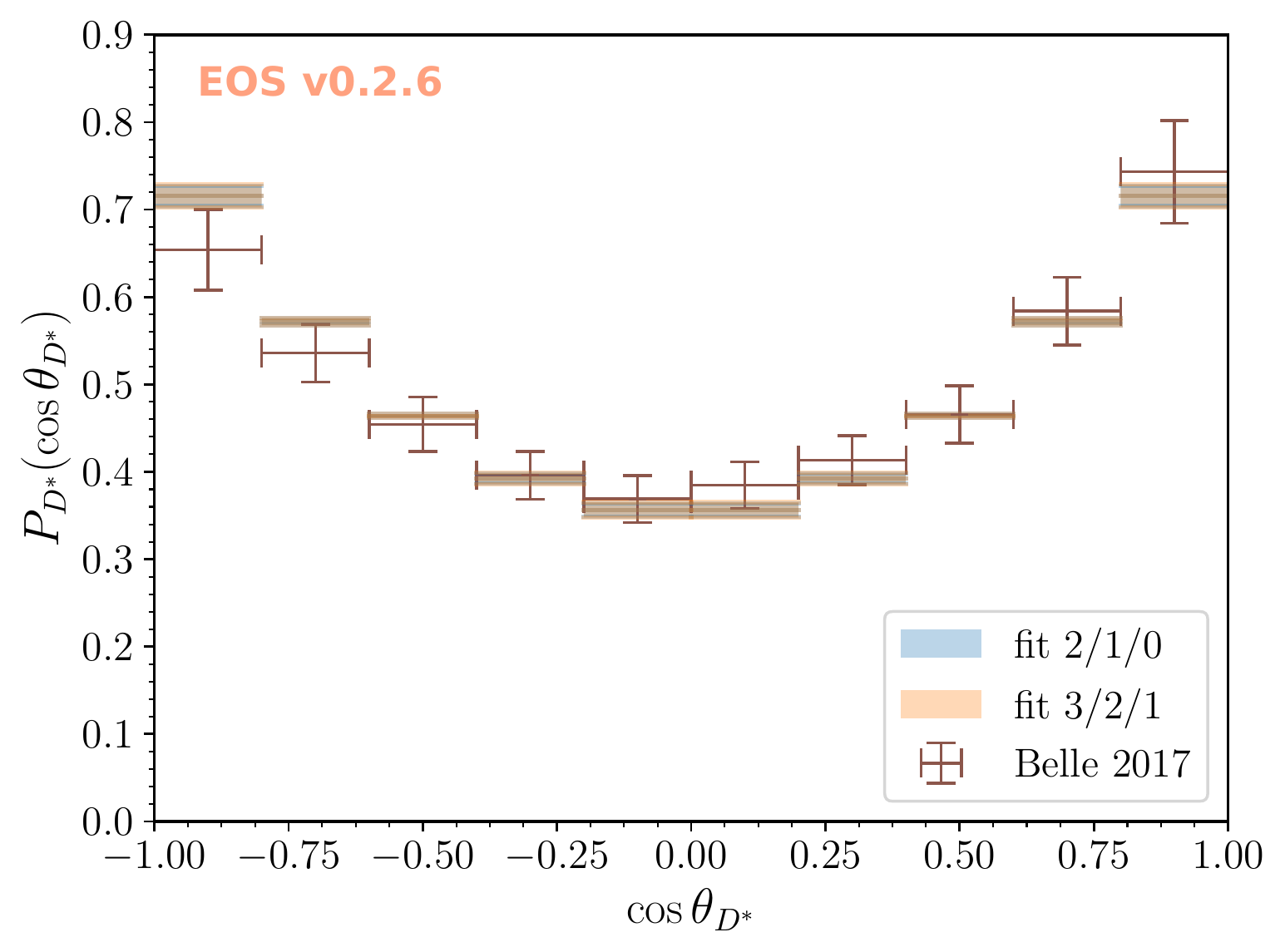} \\
        \includegraphics[width=.49\textwidth]{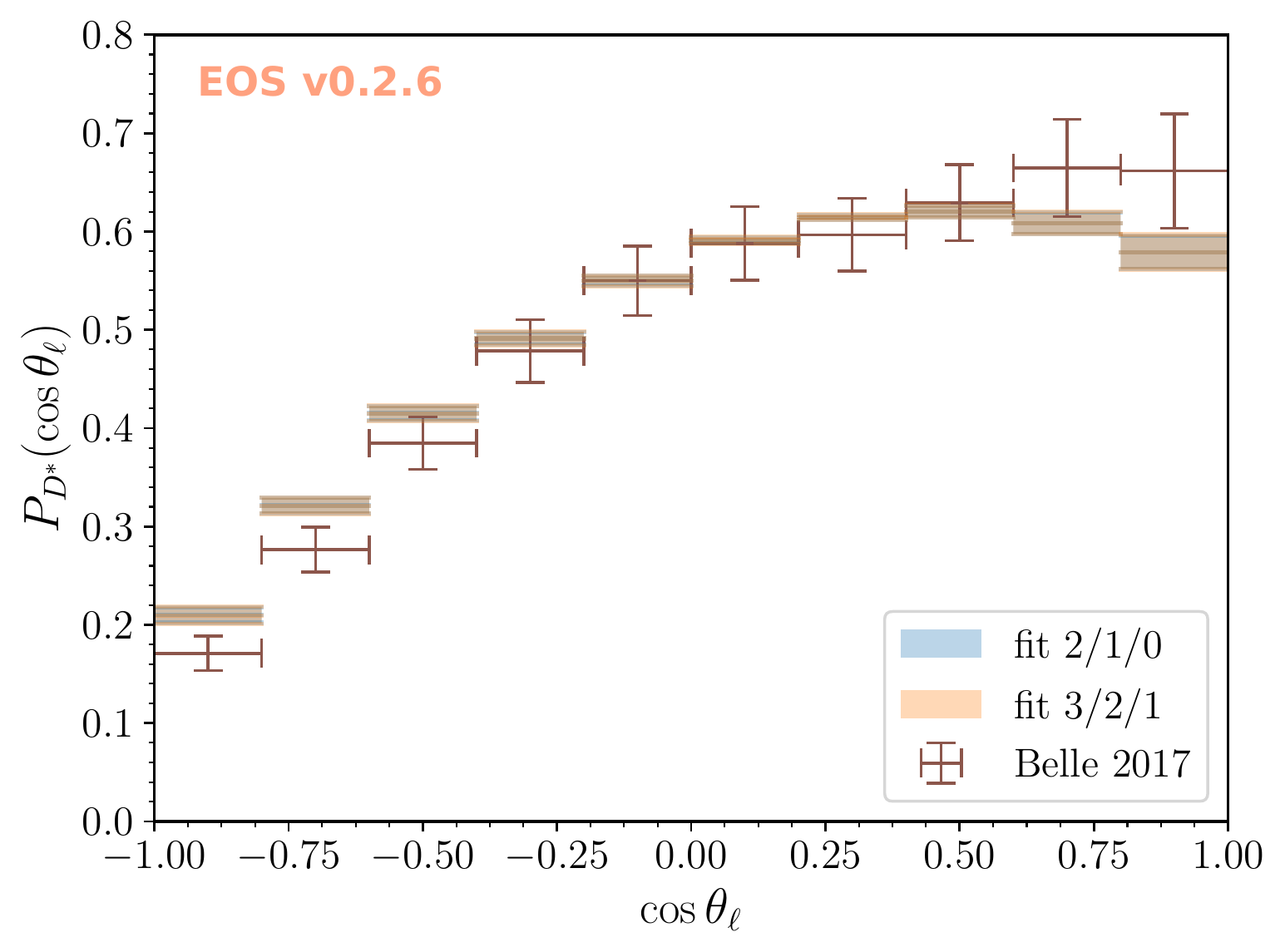} &
        \includegraphics[width=.49\textwidth]{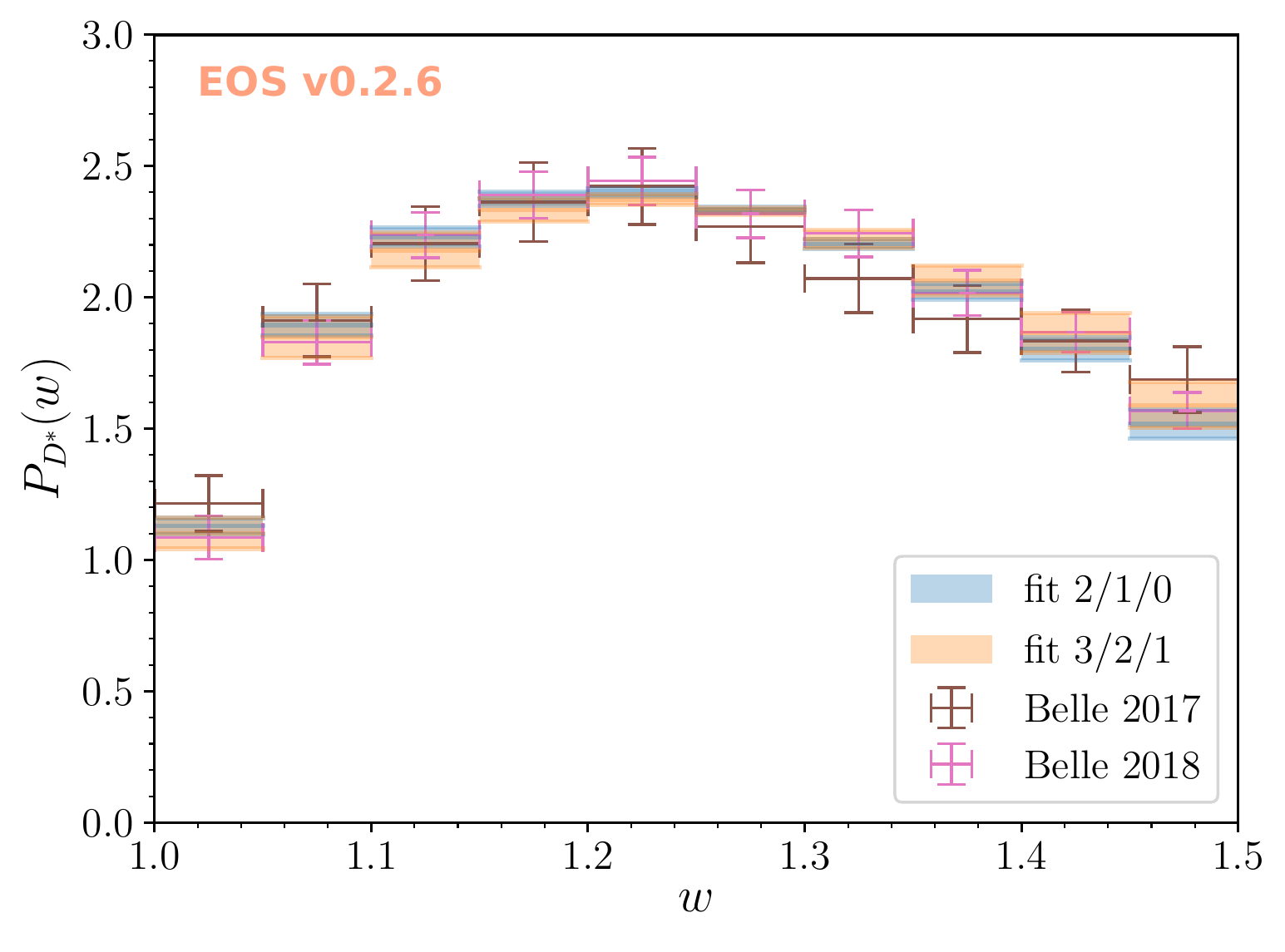}    \\
        \includegraphics[width=.49\textwidth]{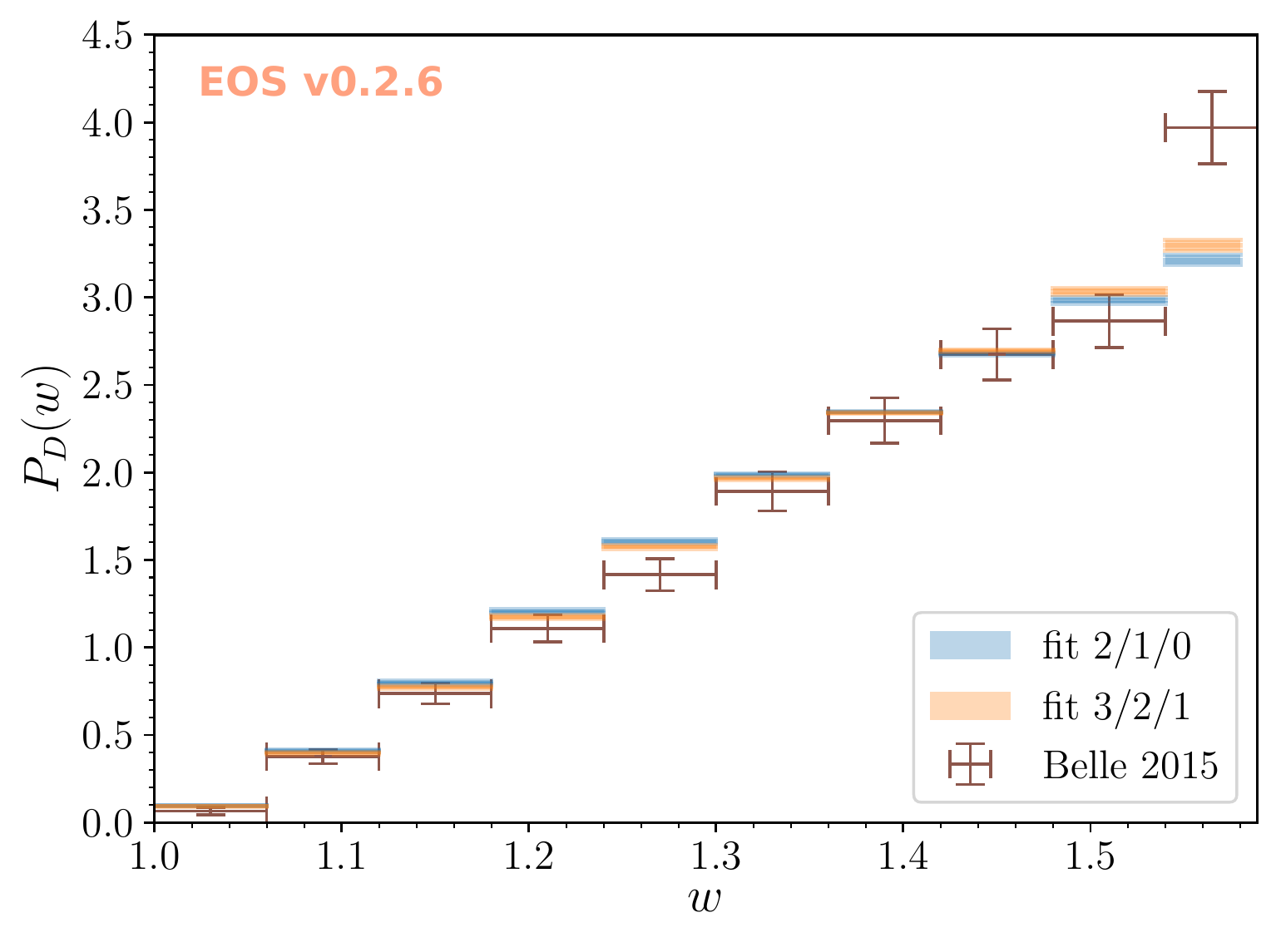}        \\
    \end{tabular}
    \caption{%
        The four 1D kinematical probability distibutions arising from the full 4D differential
        decay rate $\bar{B}\to D^*\{e^-,\mu^-\}\bar{\nu}$ next to one of the 1D kinematical distribution
        in $\bar{B}\to D\{e^-,\mu^-\}\bar\nu$ decays. We show all available constraints published by
        the Belle collaboration along side the $68\%$ probability envelopes based on the two fit models
        fitted to only theory inputs. Note that for the 2018 Belle results we have unfolded the
        results ourselves; see the text for details.
    }
    \label{fig:PDFs}
\end{figure}

In a first step, we compare in \reffig{PDFs} our posterior predictions for the kinematical PDFs with the experimental results. 
Both of our fit models yield visually indistinguishable posterior predictions for the three angular distributions
$P(\chi)$, $P(\cos\theta_\ell)$ and $P(\cos\theta_{D^*})$ in $\bar{B}\to D^*\{e^-,\mu^-\}\bar\nu$.  The agreement
between our predictions and the experimental measurements for $P(\cos\theta_\ell)$ is visibly worse than the excellent
agreement for the remaining two angular distributions. However, we find that our predictions for these three
distributions are considerably more precise than the experimental results. We therefore conclude that the latter do not
further constrain the form factor parameters within our two models; we hence abstain from using them in the following. 
However, we find that the results for the distributions $P_D(w)$ and $P_{D^*}(w)$ do have the potential to further
constrain the form factor parameters.\\

\begin{table}
    \renewcommand{\arraystretch}{1.1}
    \begin{tabular}{l c cc cc cc cc}
        \toprule
        order
            & function $f$
            & \multicolumn{2}{c}{$f(1)$}
            & \multicolumn{2}{c}{$f'(1)$}
            & \multicolumn{2}{c}{$f''(1)$}
            & \multicolumn{2}{c}{$f'''(1)$}
            \\
        \midrule
        $1/m_Q^0$
            & $\xi$
            & $+1.00$ & ---
            & $-1.14$ & $[-1.32, -0.93]$
            & $+1.88$ & $[+1.57, +2.52]$
            & $-3.29$ & $[-5.13, -2.90]$
            \\
        \midrule
        \multirow{3}{*}{$1/m_Q^1$}
            & $\hat\chi_2$
            & $-0.06$ & $[-0.08, -0.04]$
            & $-0.00$ & $[-0.02, +0.02]$
            & $+0.06$ & $[-0.21, +0.16]$
            & ---     & ---
            \\
        ~
            & $\hat\chi_3$
            & $+0.00$ & ---
            & $+0.04$ & $[+0.02, +0.06]$
            & $-0.05$ & $[-0.16, -0.04]$
            & ---     & ---
            \\
        ~
            & $\hat\eta$
            & $+0.60$ & $[+0.44, +0.79]$
            & $-0.02$ & $[-0.18, +0.18]$
            & $-0.04$ & $[-0.84, +0.32]$
            & ---     & ---
            \\
        \midrule
        \multirow{6}{*}{$1/m_Q^2$}
            & $\hat\ell_1$
            & $+0.12$ & $[-0.10, +0.36]$
            & $-5.78$ & $[-12.5, -0.16]$
            & ---     & ---
            & ---     & ---
            \\
        ~
            & $\hat\ell_2$
            & $-1.89$ & $[-2.26, -1.54]$
            & $-3.14$ & $[-9.53, +1.31]$
            & ---     & ---
            & ---     & ---
            \\
        ~
            & $\hat\ell_3$
            & $+0.86$ & $[-8.29, +5.17]$
            & $+0.06$ & $[-2.96, +9.55]$
            & ---     & ---
            & ---     & ---
            \\
        ~
            & $\hat\ell_4$
            & $-2.02$ & $[-3.53, -0.75]$
            & $-0.05$ & $[-1.88, +1.71]$
            & ---     & ---
            & ---     & ---
            \\
        ~
            & $\hat\ell_5$
            & $+3.79$ & $[+0.16, +5.20]$
            & $-1.40$ & $[-2.63, +3.26]$
            & ---     & ---
            & ---     & ---
            \\
        ~
            & $\hat\ell_6$
            & $+3.53$ & $[-0.67, +6.43]$
            & $+0.04$ & $[-3.43, +4.49]$
            & ---     & ---
            & ---     & ---
            \\
        \bottomrule
    \end{tabular}
    \renewcommand{\arraystretch}{1.0}
    \caption{Best-fit point for the parameters of the $3/2/1$ model in a simultaneous fit to theory constraints and all
        available experimental measurements. Uncertainty ranges presented here are meant for illustrative purpose only,
        and should not be interpreted a standard deviations due to non-Gaussianity of the joint posterior.
    }
    \label{tab:BFP}
\end{table}

In a second step, we fit the HQE expressions for the form factors simultaneously to the previously discussed theory constraints
and different sets of publicly available experimental results for $P_{D}(w)$ and $P_{D^*}(w)$.
These sets are: only $P_{D^*}(w)$ from the 2017 data, only $P_{D^*}(w)$ from the 2018 data, and the combination of all
experimental results for $P_{D^{(*)}}(w)$.
For all these sets we find that the simultaneous fits show excellent agreement between the theoretical constraints and
the experimental PDFs.
A summary of the goodness of fit in the best-fit points of all considered sets is presented in \reftab{gof}. Our nominal
best-fit point, obtained in the $3/2/1$ model, is presented in \reftab{BFP}.
Due to the non-Gaussianity of the posterior we refrain from providing linear correlations.
We note that the slopes of the subsubleading IW functions $\ell_i'(1)$ are all compatible with zero at $\simeq 68\%$
probability.\\

Our predictions for the $\bar B\to D^{(*)}\tau^-\bar\nu$ observables including the experimental information read:
\begin{equation}
\begin{aligned}\label{eq:bctaunupred2}
    R_{D}
        & = 0.297 \pm 0.003\,, &
    R_{D^*}
        & = 0.250 \pm 0.003\,, \\
    P_\tau^{D}
        & = 0.321 \pm 0.003\,, &
   -P_\tau^{D^*}
        & = 0.496 \pm 0.015\,, \\
        &                      &
    F_L
        & = 0.464 \pm 0.010\,.
\end{aligned}
\end{equation}
While the predictions for $\bar B\to D\tau^-\bar\nu$ remain unchanged, we find a shift of $\sim 0.5\sigma$ for the three
$\bar B\to D^*\tau^-\bar\nu$ observables. This is not surprising, given the high precision of the available $\bar B\to
D$ form factor constraints.\\ 

Finally, we produce posterior predictions for the integrated branching ratios of $\bar{B}\to D^{(*)}\ell^-\bar\nu$
decays in units of $|V_{cb}|^2$. We choose to present our results for the $\bar{B}^0$ mode only. Our results are listed
in the top half of \reftab{BRs+Vcb}. We then proceed to extract the value of $|V_{cb}|$ from the isospin averages of the
respective branching ratios. Our results for $|V_{cb}|$ are listed in the bottom half of \reftab{BRs+Vcb}. The isospin
average of the necessary branching ratios, expressed as branching ratios of the $\bar{B}^0$ mode, are:
\begin{equation}
\begin{aligned}
    \mathcal{B}(\bar{B}^0\to D^+\{e^-,\mu^-\}\bar\nu)
        & = (2.24 \pm 0.07) \%\,, &
    \mathcal{B}(\bar{B}^0\to D^{*+}\{e^-,\mu^-\}\bar\nu)
        & = (5.11 \pm 0.10) \%\,.
\end{aligned}
\end{equation}
Our nominal result for the exclusive determination of $|V_{cb}|$, obtained by combining all available theoretical and
experimental information, is:
\begin{equation}
    \left|V_{cb}^\text{excl}\right| = \left(40.3 \pm 0.8\right) \cdot 10^{-3}\,.
\end{equation}
Its agreement with the individual values from $\bar B\to D^{(*)}\ell^-\bar \nu$ is excellent. Averaging the two
exclusive determinations with the inclusive one \cite{Gambino:2016jkc}, we find $|V_{cb}|=(41.3\pm0.5)\times
10^{-3}$, where the three values are compatible at the $1.2\sigma$ level.

\begin{table}[t]
    \renewcommand{\arraystretch}{1.2}
    \begin{tabular}{l c c c c c}
    \toprule
    ~                                                        &
        \multicolumn{5}{c}{scenarios}                        \\
    model                                                    &
        $2/1/0$                                              &
        $3/2/1$                                              &
        $3/2/1$                                              &
        $3/2/1$                                              &
        $3/2/1$                                              \\
    exp.~likelihood                                          &
        ---                                                  &
        ---                                                  &
        2017                                                 &
        2018                                                 &
        all exp.                                             \\
    \midrule
    $\mathcal{B}(\bar{B}^0\to D^+\{e^-,\mu^-\}\bar\nu)/|V_{cb}|^2$   &
        $12.99 \pm 0.35$                                     &
        $13.48 \pm 0.37$                                     &
        ---                                                  &
        ---                                                  &
        $13.56 \pm 0.35$                                     \\
    $\mathcal{B}(\bar{B}^0\to D^{*+}\{e^-,\mu^-\}\bar\nu/|V_{cb}|^2$ &
        $32.33 \pm 1.28$                                     &
        $33.16 \pm 2.15$                                     &
        $31.74 \pm 1.46$                                     &
        $32.19 \pm 1.03$                                     &
        $32.00 \pm 1.03$                                     \\
    correlation                                              &
        $0.34$                                               &
        $0.14$                                               &
        ---                                                  &
        ---                                                  &
        $0.10$                                               \\
    \midrule
    $|V_{cb}|\times 10^3$ from $\bar{B}\to D\{e^-,\mu^-\}\bar\nu$   &
        $41.5 \pm 1.2$                                       &
        $40.7 \pm 1.2$                                       &
        ---                                                  &
        ---                                                  &
        $40.6 \pm 1.1$                                       \\
    \multirow{2}{*}{%
    $|V_{cb}|\times 10^3$ from $\bar{B}\to D^*\{e^-,\mu^-\}\bar\nu$}&
        $39.8 \pm 1.2$                                       &
        $39.3 \pm 1.7$                                       &
        $40.1 \pm 1.3$                                       &
        $39.8 \pm 1.0$                                       &
        $40.0 \pm 1.1$                                       \\
                                                             &
        ---                                                  &
        ---                                                  &
        $(39.5\pm1.9)$                                       &
        $(39.0\pm1.3)$                                       &
        ---                                                  \\
    $|V_{cb}|\times 10^3$ combined incl. corr.               &
        $40.7 \pm 1.0$                                       &
        $40.2 \pm 1.0$                                       &
        ---                                                  &
        ---                                                  &
        $40.3 \pm 0.8$                                       \\
    \bottomrule
    \end{tabular}
    \renewcommand{\arraystretch}{1}
    \caption{%
        Posterior predictions for the branching ratios of $\bar{B}^0\to \lbrace D^+,D^{*+}\rbrace \{e^-,\mu^-\}\bar\nu$
        decays in units of $|V_{cb}|^{2}$, as well as the values of $|V_{cb}|$ extracted from the isospin-averaged
        branching ratios. Throughout we use the HFLAV world averages \cite{Amhis:2016xyh} for the determination
        of $|V_{cb}|$. For the columns marked $2017$ and $2018$, the values in parentheses are obtained by
        using only the respective Belle measurements of the branching ratios
        \cite{Abdesselam:2017kjf,Abdesselam:2018nnh}.
        The row for the \emph{combined} $|V_{cb}|$ takes into account correlations in the posterior predictions, but not
        the small and unpublished correlations in the HFLAV world averages.
    }
    \label{tab:BRs+Vcb}
\end{table}

\section{Summary and outlook}
\label{sec:summary}

In this work we carry out a comprehensive analysis of the full set of $\bar B\to D^{(*)}\ell^-\bar\nu$ form factors. The
basis of our analysis is the Heavy-Quark Expansion (HQE) up to $\mathcal O(\eps^2)$, where our power-counting is
defined as $\eps_b\sim\eps_c^2\sim\alpha_s/\pi\sim\eps^2$. By determining the coefficients of this
expansion from all available theoretical constraints we are able to predict the full set of form factors with high
precision. This allows for their consistent and accurate use in a variety of phenomenological applications
without the assumption of absence of NP effects in $b\to c$ transitions with light leptons. Our work focuses on two
applications: precision predictions for $\bar B\to D^{(*)}\tau^-\bar\nu$ observables and accurate determinations of
$|V_{cb}|$ from $\bar B\to D^{(*)}\lbrace e^-,\mu^-\rbrace \bar\nu$ decays.\\

We find excellent agreement between the various theoretical constraints on the relevant hadronic matrix elements. The
minimal viable fit model is found to be the $2/1/0$ model, where the numbers refer to the order in the $z$ expansion of
the leading, subleading, and subsubleading Isgur-Wise (IW) functions, respectively. To account for systematic uncertainties
inherent to the HQE, we increase the order of the $z$ expansion for all three sets of IW functions, which defines
our nominal fit model. Within our analysis we pay particular attention to the subsubleading terms in the HQE. In
previous analyses it turned out to be necessary to include at least two such terms. Our analysis is the first to include
the full set of IW functions at the order $\mathcal O(1/m_c^2)$. We find that the expansion in $1/m_Q$ is well-behaved,
similar to what has been found in a recent analysis of $\Lambda_b\to \Lambda_c$ form factors \cite{Bernlochner:2018kxh}.
Based on these findings we expect the terms at $\mathcal O(\eps^3)$ to be negligible at the present level of precision.
This assumption should be revisited once more precise theoretical and experimental information becomes available.\\

Our predictions for $\bar B\to D^{(*)}\tau^-\bar\nu$ observables benefit from the improved treatment of the HQE. This
is reflected by significantly smaller uncertainties compared to previous analyses, while staying compatible at the
$1\sigma$ level. Predictions with and without the use of experimental inputs are given in \refeq{bctaunupred2}
and~\refeq{bctaunupred}, respectively.\\

Our determinations of $|V_{cb}|$ from $D$ and $D^*$ final states are mutually compatible and also
compatible with the inclusive determination at the $1.2\sigma$ level. Unlike for CLN analyses, we find no tension with
the BGL determinations. Our nominal result for $|V_{cb}|$ using all exclusive experimental inputs reads
\begin{equation*}
    \left|V_{cb}^\text{excl}\right| = \left(40.3 \pm 0.8\right) \cdot 10^{-3}\,.
\end{equation*}

The upcoming lattice analyses of four of the $\bar B\to D^*$ form factors at nonzero recoil
\cite{Vaquero:2019ary,Kaneko:2018mcr,Bhattacharya:2018ibo} will benefit our approach and help to
determine the HQE parameters to even higher precision. 

\acknowledgments

We are grateful to Florian Bernlochner and Phillip Urquijo for helpful discussions regarding the Belle data.\\

The work of MB is supported by the Deutsche Forschungsgemeinschaft (DFG, German Research Foundation) under grant  396021762 - TRR 257.
The work of MJ is supported by the the DFG Excellence Cluster ``Origins and Structure of the Universe'' and the
Italian Ministry of Research (MIUR) under grant PRIN 20172LNEEZ.
The work of DvD is supported by the DFG within the Emmy Noether Programme under grant DY130/1-1 and
the DFG Collaborative Research Center 110 “Symmetries and the Emergence of Structure in QCD”. We thank the DFG
Excellence Cluster ``Origins and Structure of the Universe'' for supporting a short-term visit of MB.
We are grateful to both the Munich Institute for Astro- and Particle Physics and the Institute for Nuclear Theory at the
University of Washington for their hospitality during parts of this work.

\appendix

\section{Statistical Treatment of the Unitarity Bounds}
\label{app:bounds}

We consider a positive semi-definite function $B(\vec{\theta}\,)$, where $\vec{\theta}$ denotes the parameters of interest. In the context
of our work, $B$ corresponds to any of the previously discussed strong unitarity bounds.
Moreover, we have $B(\vec{\theta}\,) \leq \chi_B$. The probability density of the parameters $\vec{\theta}$ can then be expressed as
\begin{equation}
    P(\vec{\theta}\,) = \frac{N(\vec{\theta}\,)}{\int d\vec{\theta}\, N(\vec{\theta}\,)}\,,
\end{equation}
where we use
\begin{equation}
    N(\vec{\theta}\,) = \int_{-\infty}^{+\infty} d\chi_B\,\mathbbm{1}\left[B(\vec{\theta}\,) \leq \chi_B\right]\, P(\chi_B)\,.
\end{equation}
In the above $\mathbbm{1}$ denotes the indicator function, and $P(\chi_B)$ is the probability density for the upper bound on $B$.
For the case that $\chi_B$ is precisely known, $P(\chi_B)$ can be approximated as a Dirac delta and consequently $P(\vec{\theta}\,)$
is given by the indicator function. For our application, the upper bounds carry a significant theoretical uncertainty, which
we model as independent Gaussian distribution for each bound, centered around a value $\mu_B$ and with a standard deviation of
$\sigma_B$. In that case we obtain:
\begin{equation}
    -2 \ln N(\vec{\theta}\,) \simeq \begin{cases}
        0                                                               & B(\vec{\theta}\,) \leq \mu_B\\
        \left(\frac{B(\vec{\theta}\,) - \mu_B}{\sigma_B}\right)^2       & \text{otherwise}
    \end{cases}\,.
\end{equation}

\section{Form Factor Definitions}

\subsection{$\bar{B}\to D$}

The $\bar{B}\to D$ form factors have been defined in Eqs.~\eqref{eq:BPVFF}-\eqref{eq:BPTFF}. Their translation to the heavy-quark matrix elements defined in Ref.~\cite{Bernlochner:2017jka} reads
\begin{align}
    f_+ & = \frac{1}{2\sqrt{r}}\,\left[(1 + r) h_+ - (1 - r) h_-\right]\,,\\
    f_- & = \frac{1}{2\sqrt{r}}\,\left[(1 + r) h_- - (1 - r) h_+\right]\,,\\
    f_0 & = f_+ + \frac{1 + r^2 - 2 r w}{1 - r^2}\, f_-\,,\\
    f_T & = \frac{1 + r}{2 \sqrt{r}}\, h_T\,,
\end{align}
where $r \equiv M_D / M_B$.

\subsection{$\bar{B}\to D^*$}

The form factors for $\bar{B}\to D^*$ are defined in Eqs.~\eqref{eq:BVVFF}-\eqref{eq:BVTFF}.
The factors in Eq.~\eqref{eq:BVTFF} have been chosen such that the conventions of Ref.~\cite{Ball:2004rg} for the following
contractions are recovered:\footnote{Note again the different sign convention for the Levi-Civita tensor used there.}
\begin{align}
    \bra{D^*(k, \eta)} \bar{c} q^\nu\sigma_{\mu\nu} b \ket{\bar B(p)}
        & = -2 i \epsilon_{\mu\nu\rho\sigma} \eta^{*\nu}p^\rho k^\sigma T_1(q^2)\,,\\
    \bra{D^*(k, \eta)} \bar{c} q^\nu\sigma_{\mu\nu}\gamma_5 b \ket{\bar B(p)}
        & = \eta^*_\alpha \left\{\left[(M_B^2-M_{D^*}^2)g^{\alpha}_{\mu}-p^\alpha (p+k)_\mu\right]T_2(q^2)+p^\alpha
        \left[q_\mu-\frac{q^2}{M_B^2-M_{D^*}^2}(p+k)_\mu\right]T_3(q^2)\right\}\,.
\end{align}

Translating again to the heavy-quark form factors in \cite{Bernlochner:2017jka}, we find\footnote{Note the different convention for the behaviour under time reversal in that article, which necessitates a factor of 'i' in the comparison on the side of the QCD form factors.}
\begin{align}
    V
        & = \frac{1 + r}{2 \sqrt{r}} h_V\,,\\
    A_0
        & = \frac{1}{2 \sqrt{r}} \left[ (w + 1) h_{A_1} + (r w - 1) h_{A_2} + (r - w) h_{A_3}\right]\,,\\
    A_1
        & = \frac{\sqrt{r}\,(1 + w)}{1 + r} h_{A_1}\,,\\
    A_{12}
        & = \frac{1 + w}{8 \sqrt{r}}\left[(w - r) h_{A_1} - (w - 1) r h_{A_2} - (w - 1) h_{A_3}\right]\,,\\
    T_1
        & = -\frac{1}{2\sqrt{r}}\left[(1 - r) h_{T_2} - (1 + r) h_{T_1}\right]\,,\\
    T_2
        & = \frac{1}{2\sqrt{r}}\left[\frac{2 r (w + 1)}{1 + r} h_{T_1}-
            \frac{2r(w - 1)}{1 - r} h_{T_2}\right]\,,\\
    T_3
        & = \frac{1}{2\sqrt{r}}\left[(1 - r) h_{T_1} - (1 + r) h_{T_2} + \left(1 - r^2\right) h_{T_3} \right]\,,
\end{align}
where $r \equiv M_{D^*} / M_B$.

\subsection{Details on the HQE for the form factors}
\label{app:FF:HQE}

A generic heavy-quark form factor $h$ is expanded in $\alpha_s/\pi$, $\eps_Q \equiv \bar{\Lambda}/2m_Q$, and $z$ as follows: 
\begin{equation}
    h(z) = \xi(w)\hat h(w) = \xi(w)\left(
        a + \frac{\alpha_s}{\pi}b\left[\hat C_i(w)\right]
        + \eps_b \, c_b\left[\hat L_i(w)\right]
        + \eps_c \, c_c\left[\hat L_i(w)\right]
        + \eps_c^2 \, d\left[\hat\ell_i(w)\right]
    \right)_{w = w(z)}\,
\end{equation}
where $a=1,0$, depending on whether the form factor vanishes in heavy-quark limit or is proportional to the leading IW
function $\xi(w)$, and $b$, $c_b$, $c_c$, and $d$ represent linear functionals of their arguments.  The $\alpha_s$
corrections are written in terms of the leading order results for the HQET Wilson coefficients $\hat C_i(w)$; see
\cite{Neubert:1993mb} for a review. The $\order{\eps_Q}$ corrections fulfill equations of motion that reduce the number
of independent functions $\hat{L}_i$ from six to four. These four are commonly denoted as $\hat\chi_{1,2,3}(w)$ and
$\hat\eta(w)$, where heavy-quark
spin symmetry allows $\hat \chi_1(w)$ to be absorbed into $\xi(w)$. Hence, only three independent functions remain
\cite{Neubert:1993mb}. Similarly, equations of motion relate the functions at $\order{\eps_Q^2}$ with each other. However, the set
of independent functions is comprised by more than six functions \cite{Falk:1992wt}. Since we are only interested in the $\eps_c^2$
corrections, it suffices to define the first six IW functions at $\order{\eps_Q^2}$ as the six functions arising at
$\order{\eps_c^2}$.  Finally, all functions are expanded in $z$ according to \refeq{def:w-expansion}, to the order
given in the corresponding model.

\bibliographystyle{apsrev4-1}
\bibliography{references}

\end{document}